\documentclass[article]{./jss}

\usepackage{thumbpdf}

\author{J~Antonio~Rivero~Ostoic\\ Aarhus University}
\title{Algebraic analysis of multiple social networks\\ with \pkg{multiplex}}

\Plainauthor{J~Antonio~Rivero~Ostoic} 
\Plaintitle{Algebraic analysis of multiple social networks with multiplex} 
\Shorttitle{\pkg{multiplex}: Algebraic analysis of multiple social networks} 

\Abstract{
\pkg{multiplex} is a computer program that provides algebraic tools for the analysis of multiple network structures within the \proglang{R} environment. Apart from the possibility to create and manipulate multivariate data representing multiplex, signed, and two-mode networks, this package offers a collection of functions that deal with algebraic systems ---such as the partially ordered semigroup, and balance or cluster semirings--- their decomposition, and the enumeration of bundle patterns occurring at different levels of the network. Moreover, through Galois derivations between families of the pairs of subsets in different domains it is possible to analyze affiliation networks with an algebraic approach. Visualization of multigraphs, different forms of bipartite graphs, inclusion lattices, Cayley graphs is supported as well with related packages.
}
\Keywords{social network analysis, relational algebra, graph visualization, \proglang{R}}
\Plainkeywords{social network analysis, relational algebra, graph visualization, R} 

\Address{
  Antonio Rivero Ostoic\\
  School of Culture and Society\\
  Aarhus University\\
  E-mail: \email{multiplex@post.com}\\
  URL: \url{https://github.com/mplex}\\
}

\usepackage{url}
\usepackage{amsmath}
\usepackage{amssymb}
\usepackage{amsfonts}
\usepackage{lmodern}
\usepackage[singlespacing]{setspace}
\newcommand{\Xx}{\mathbf{X}}
\newcommand{\Rr}{\mathbf{R}}
\newcommand{\Aa}{\mathbf{A}}

\newcommand{\cC}{\footnotesize\textsf{C}\normalsize}%
\newcommand{\fF}{\footnotesize\textsf{F}\normalsize}%
\newcommand{\kK}{\footnotesize\textsf{K}\normalsize}%
\newcommand{\aA}{\small\textsf{A}\normalsize}%
\newcommand{\nN}{\small\textsf{N}\normalsize}%
\newcommand{\mM}{\small\textsf{M}\normalsize}%

\begin{document}




\section[Introduction]{Introduction}
Social networks are sets of collective relations between different actors and the discipline that studies these types of systems is called ``social network analysis'' \citep{WassFau1994}. Multiple ---hereafter multiplex--- networks are special types of social systems where actors are connected at several levels such as people who have simultaneously business collaboration relations and informal friendship ties, or organizations that cooperate and compete at the same time, etc. Such arrangements are inherently complex due to the ``relationships between relations'', which represents a higher level of abstraction in the social structure, and we need distinctive methods to preserve the multiplicity of these ties in the analysis.

A fruitful methodological framework for the analysis of multiplex networks is provided by \emph{relational algebra} \citep{Pattison1993, DegFor1999} with a number of algebraic objects that are able to represent different types of complex structures a defined social milieu. The algebraic representation of these particular systems serves to uncover their ``relational interlock'', which is represented by different types of algebraic constraints, and that allows making a substantial interpretation of the network structure. In this paper, particular types of complex systems studied within an algebraic approach are multiplex networks with different kinds of positively valued relations, signed networks with ties having opposite valences, and affiliation networks that are arrangements with two domains, one for the actors and the other for events or categories.

Apart from the well-known computer programs for the analysis of social networks like Ucinet \citep{ucinet:02}, Pajek \citep{pajek:98}, and PNet \citep{pnet:09}, there is a number of packages for making diverse types of social network analyzes within the \proglang{R} environment \citep{RCT:2015}. Notably \pkg{sna} \citep[]{Butts08, sna:16} and \pkg{igraph} \citep{igraph:15} are popular within the social network analysis community, not only for measuring structural indices of equivalence or distance, but also for the visualization of graphs representing the network structure. A more strictly statistical approach for the analysis of network data is found in statnet \pkg{statnet} \citep{statnet:16} and its related packages, particularly \pkg{ergm} \citep{ergm:16} that can simulate and fit networks based on exponential random graph models, whereas \pkg{RSiena} \citep{RSiena:13} is the \proglang{R} implementation of the Stochastic Actor Oriented Model \citep{Snijders01} to longitudinal network data. Other programs that have focus on multiple network structures are MuxViz \citep{Muxviz:17}, Multidimensional Network Analysis \citep{MNA:17}, and Multilayer Networks Library \citep{Pymnet:17}. The former combines \proglang{R} with \proglang{GNU Octave}, whereas the other two platforms are \proglang{Java} and \proglang{Python} libraries respectively.

One of the first computer programs to perform algebraic analyzes of social networks is found in ASNET, which is a suite of modules latter incorporated in PACNET \citep{PattEtAl:00, Asnet:95}. Although one can definitely find several other packages to make specific types of analysis of network data in \proglang{R},\footnote{See the \emph{CRAN Task View: Statistics for the Social Sciences} for more examples.} at this time there are any active programs that focus specifically on multiplex network structures with algebraic approaches. In this sense, \pkg{multiplex} \citep{multiplex:19} is a computer package that delivers a collection of functions with algebraic procedures for the analysis of various kinds of multiplex networks within the \proglang{R} setting. Among other things, this package combines algebraic systems like the partially ordered semigroup or the semiring structure together with the relational bundles occurring in different types of multivariate network data sets. There is also an algebraic approach for dealing with two-mode networks that is made through Galois derivations between families of the pair of subsets. Moreover, in conjunction with \pkg{multigraph} \citep{multigraph:19}, it is possible to visualize multiplex networks in the form of ``multigraphs'' or graphs having parallel edges, Cayley graphs and much more with the graphical possibilities of the built-in \proglang{R} libraries, particularly \pkg{graphics}.

\section[Multiplex social networks]{Multiplex social networks}
Humans are collective actors that are related to each other within a defined social system at different levels. The multiplicity inherent to the social relations makes the social system a complex structure that is operationalized through the concept of ``multiplex social network'' \citep{BoorWhite1976, BrePat86, LazPatt99} for the analysis.
In formal terms, a \emph{social network} $X$ is a setting of $n$ social entities $N=\{i \mid i$  is an entity, for  $i=1,2,\dots, n\}$ measured under a collection of social ties $R = \{ ( i, j ) \mid i$  `has a tie to'  $j \}$ where for individuals $i,j \in X$, $X_R(i,j)=1$ represents a tie $R$ , and $X_R(i,j)=0$ denotes the lack of a tie between them. A network is directed when $(i,j)$ is an ordered pair, and the pairs on $X_R$ are usually recorded in an \emph{adjacency matrix} $A$ with size $n \times n$. 

A \emph{multiplex network} $\Xx$ combines a collection $\Rr$ of $r$ different kinds of relations, $\Rr = \{R_1, R_2, \dots, R_r\}$ measured over $N$. Each relational type is stored in a separate adjacency matrix $A_1, A_2, \dots, A_r$, which are stacked together into a single array $\Aa$ with size $n \times n \times r$. This object resembles an adjacency tensor of a third order with $n$ horizontal and vertical slices, and $r$ frontal slices.

\citet{WassFau1994} provide a comprehensive set of methods for the study of social networks including multiplex structures, whereas \citet{Pattison1993} makes an exhaustive treatment of algebraic approaches to analyze both complete and local networks, the latter are structures emerging from individual actors. A more specific review on multiplex networks and other types of complex structures is found in \citet{Kivela14}, whereas \citet{DeDomenico13} provide mathematical foundations of multilayer structures that is applicable to multiplex networks as well.

\subsubsection{Multiplex network structure}
To illustrate some algebraic analyzes, we take an example of a multiple network structure named ``Incubator C''. The object below called \code{incubC} (available in \pkg{multiplex}) represents empirically collected social relations of different types between actors in a closed setting (cf. \citet{Ostoic13} for details).
The \pkg{utils} function \code{str} display the structure of this object and we can see that there are different components that follow the metacharacter \code{$}. The arrays of adjacency matrices of individual relations are recorded in \code{net}, whereas vector \code{atnet} indicates those arrays that represent attributes. Component \code{IM} allocates arrays corresponding to \emph{image matrices} of aggregated role relations in \code{net}.

\begin{Schunk}
\begin{Sinput}
R> library("multiplex")
R> data("incC")
R> str(incC)
\end{Sinput}
\begin{Soutput}
List of 5                                              
 $ net  : num [1:22, 1:22, 1:5] 0 0 0 0 0 0 0 0 0 0 ...
  ..- attr(*, "dimnames")=List of 3                    
  .. ..$ : chr [1:22] "327" "328" "331" "339" ...      
  .. ..$ : chr [1:22] "327" "328" "331" "339" ...      
  .. ..$ : chr [1:5] "C" "F" "K" "A" ...               
 $ atnet:List of 1                                     
  ..$ : num [1:5] 0 0 0 1 1                            
 $ IM   : num [1:3, 1:3, 1:8] 1 1 1 1 1 0 1 0 1 1 ...  
...
\end{Soutput}
\end{Schunk}

The initial focus is on component \code{net}, which represents a network made of 22 actors who are linked by social ties. The 5 matrices recorded are named \code{C}, \code{F}, \code{K}, \code{A} (and \code{B}), where the last two letters correspond to two different types of actor attributes.

\begin{figure*}[t]
\vspace{-75pt}\centering
\begin{tabular}{c}
\begin{minipage}[t]{1\linewidth}
    \includegraphics[width=15cm,page=1]{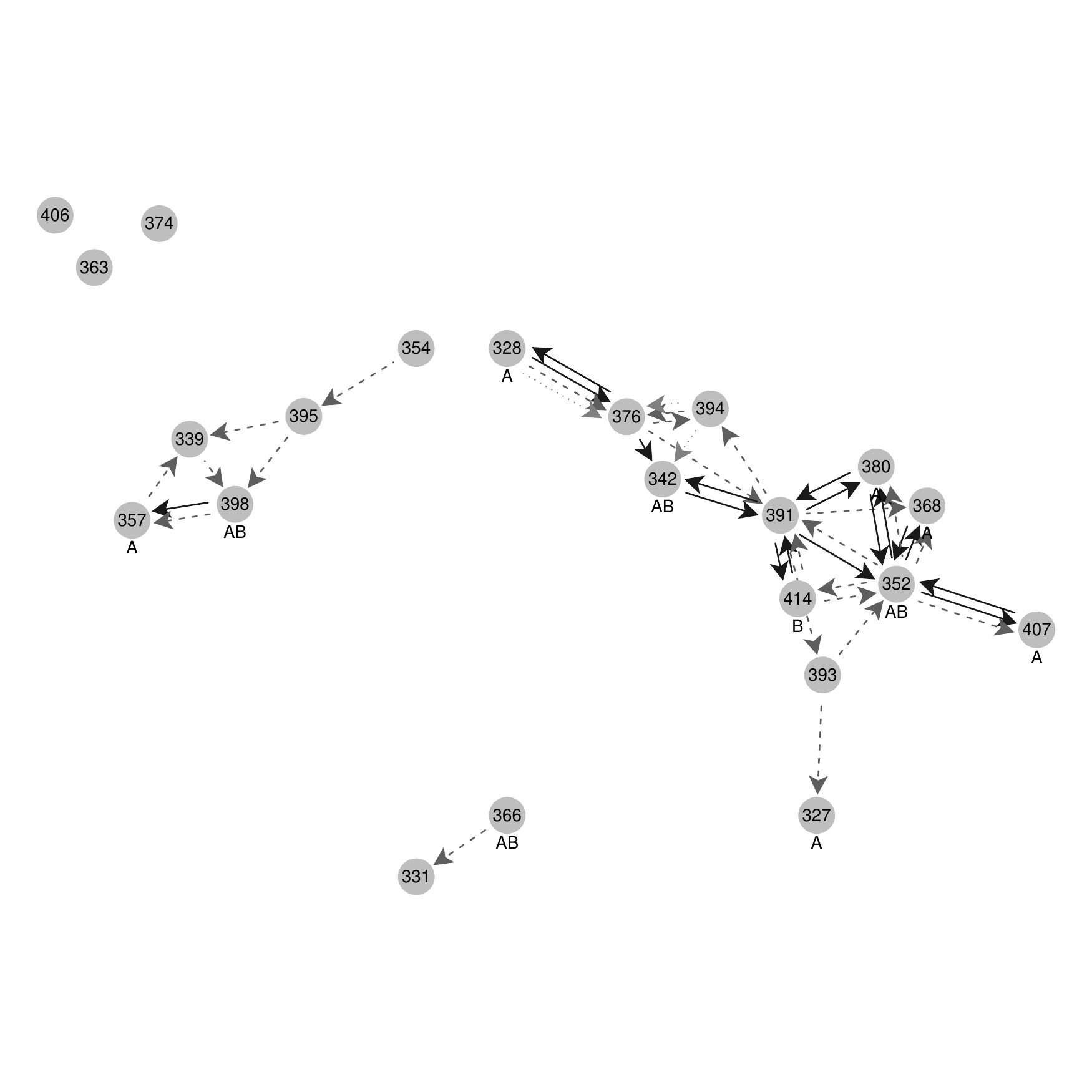}
\end{minipage}
\end{tabular}
\setlength{\abovecaptionskip}{-65pt}
  \caption{Multigraph of network ``Incubator C''.}
\label{fig:incC}
\end{figure*}

The whole network is depicted in Figure~\ref{fig:incC} as a graph with parallel edges with a \emph{force directed} algorithm \citep{Eads84, FrchRein91} implemented in \pkg{multigraph} for the visualization. The function \code{multigraph} allows us to specify a number of arguments within \proglang{R} both for nodes and for different types of edges representing actors and ties, respectively.

\begin{Sinput}
R> library("multigraph")
R> scp <- list(cex = 3, vcol = 8, bwd = .5, pos = 0, fsize = 7)
R> multigraph(incC, layout = "force", seed = 123, scope = scp)
\end{Sinput}

\subsection{Link generalizations}
Besides the ``social structure'', which is a system of relations between the actors in the network, a significant characteristic of multiplex networks is that there is also the structure produced by the relations among the relationships themselves. This kind of arrangement constitutes the network ``relational structure'' \citep{Pattison1993} and represents the intertwining of the different types of ties occurring in the system. Because of the complexity inherent in this type of arrangement, the analysis and substantial interpretation of the network relational structure and the social system in particular often constitute a great challenge.

Certainly, one approach to deal with such complexity is to reduce the network structure, and there are diverse techniques within the field of blockmodeling that produce classes of structurally related actors. However, most of the blockmodeling methods are based on the actors' embeddedness in the network, which is typically grounded on a single type of relationship \citep[see][for significant examples]{LorrWht71, DorBatFer2004}. For multiplex networks, there is a significant loss of structural information in case the condensed system fails to reflect the multiplicity of the ties. Hence, it is desirable to produce a simpler and single structure that integrates the different types of relations, and at the same time provides valuable insights of the whole system for its substantial interpretation.

Integrating relations in multiplex networks implies a form of \emph{generalization} of the links between pairs of actors in the system, and we recognize the multiplicity of the ties in the social structure either at the different levels in the relationship or by considering chains of several kinds of relations that influence the structuring process of the social system. The former is regarded as a ``wide'' extension of the pairwise link, whereas the latter generalization constitutes a ``long'' extension of the link.

The wide extension of the ties is expressed in different classes of configurations that occur at the dyadic level in the network structure and we call them ``bundles''. Such categories include the well-known patterns for single networks like the null, asymmetric, and reciprocal dyads \citep{Hol-Lei76}, but also other arrangements that have the multiplexity property, which generalize these kinds of dyadic relations. The entrainment and the exchange of ties in directed networks are fundamental patterns where at least two types of tie are involved and, although there are bundle configurations where these patterns are mixed as well, these two configurations represent entirely different realities with consequences in the structural analyzes of social phenomena in multiplex network structures.

With respect to the long extension of the link, this form of generalization corresponds to the interrelations among the ties that produce chains and paths of relations. In this sense, simple ties such as social interactions, flow of information, co-occurrence, etc. constitute primitive relations in the system whereas the concatenation of these create compound relations. ``Strings'' is a generic name for both primitives and compounds, and string relations are known metaphorically as ``words'' composed by primitives or letters from an alphabet. Even though it is likely to have an infinite number of string relations, most of them will connect precisely the same individuals and as a result, there are a limited number of isomorphic strings in the closed system that stands for the network relational structure.

\subsection[Relational Bundles]{Relational bundles}
For multiplex network structures, the wide component of the link is expressed in different types of \emph{bundle patterns}, which are configurations at the dyadic level that encode the simultaneity of the various kinds of tie. The entrainment and the exchange of ties represent two fundamental classes with the multiplexity property, and these bundle patterns are extensions of the asymmetric and reciprocal dyad, respectively. 

Formally, these classes are defined for an ordered pair of actors $(i, j) \in R_r$ as:
\begin{equation*}\label{eq:tent}
\begin{split}
\textnormal{\rm Tie Entrainment}\colon  &  \; (i,j) \in R_{1,\dots, s} \; \wedge \; (j,i) \notin R_{1,\dots, s}, \quad \text{for } 1 < s \leqslant r 
\end{split}
\end{equation*}
\vspace{-20pt}%
\begin{equation*}\label{eq:txch}
\begin{split}
\textnormal{\rm Tie Exchange}\colon     &  \;\quad (i,j) \in R_p \; \wedge \; (j,i) \in R_q,  \qquad\qquad  \text{for }  R_p \ne R_q 
\end{split}
\end{equation*}

Hence, while tie entrainment has an asymmetric character, the exchange of ties of different type implies --as the reciprocal dyad-- that the bundle pattern has a mutual character. Besides, the mixture of an asymmetric and a reciprocal dyad represents another bundle class presumably with a mutual character, but that depends on the relational content of the ties. The definition of the asymmetric and the reciprocal dyads in this case implies that the remaining levels in the relationship (if any) are null. 

To perform the bundle census or the enumeration of the bundle patterns occurring in multiplex networks, \pkg{multiplex} has function \code{bundle.census}, and we apply it to object \code{netC} that records the social relations in component \code{net} of \code{incC}.

\begin{Schunk}
\begin{Sinput}
R> netC <- incC$net[ , , 1:3]
R> bundle.census(netC)
\end{Sinput}
\begin{Soutput}[fontsize=\small]
      BUNDLES NULL ASYMM RECIP T.ENTR T.EXCH MIXED FULL
TOTAL      25  206    14     3      1      1     6    0
\end{Soutput}
\end{Schunk}

\begin{figure*}[t]
    \centering\includegraphics[width=15.5cm]{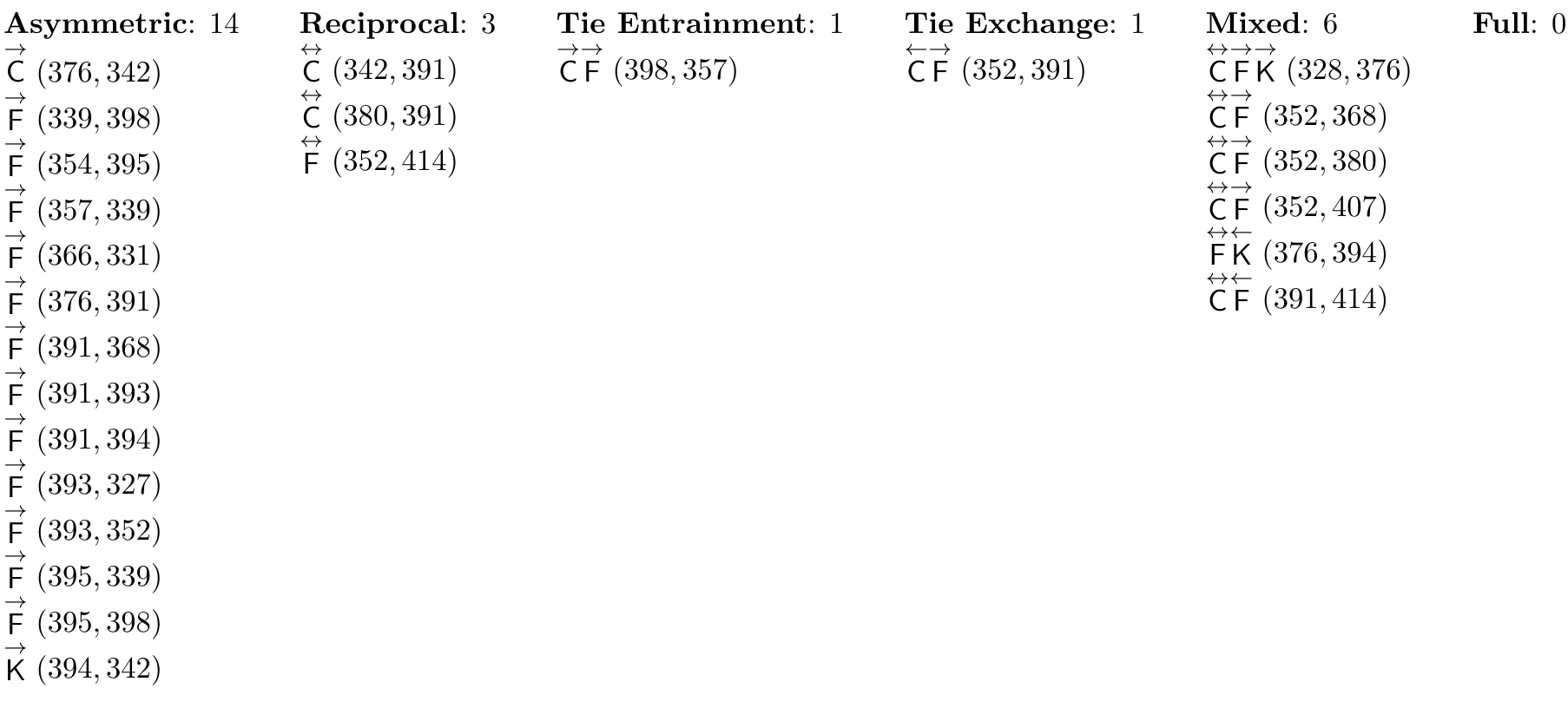}
\setlength{\abovecaptionskip}{-10pt}
  \caption{Bundle classes occurring in Incubator C.}
\label{fig:bundles}
\end{figure*}

Likewise, functions \code{bundles} and \code{summaryBundles} provide a more detailed information of the bundle patterns that exist in the network. The output from the code below produces a picture given in Figure~\ref{fig:bundles} where the dimnames attribute of each array in \code{netC} (i.e., \cC, \fF, and \kK) serves to denote the type of tie involved in the bundle with an arc on the top indicating the direction of the tie between the actors.

\begin{Schunk}
\begin{Sinput}
R> summaryBundles(bundles(netC), latex = TRUE, file = "./path")
\end{Sinput}
\end{Schunk}

\subsubsection{Relational systems of bundle classes}
The information provided by the different bundle classes allows us to evaluate the characteristic structure of the multiplex network, which is the basis for distinguishing particular ``systems'' inside the network for the developing of specific theories. For this, the \code{bonds} argument in function \code{rel.sys} allows us to disentangle bundle patterns or classes of them that are ``bonds''. For instance, we identify just the tie exchange bundle occurring in \code{netC} (cf. Fig.~\ref{fig:incC}), which is recorded in component \code{Ties} of the output.

\begin{Schunk}
\begin{Sinput}
R> rel.sys(netC, type = "tolist", bonds = "txch")$Ties
\end{Sinput}
\begin{Soutput}[fontsize=\small]
$C
[1] "391, 352"

$F
[1] "352, 391"

$K
character(0)
\end{Soutput}
\end{Schunk}

Because of its compactness, each type of relation is characterized by a vector with a ``list of pairwise relations'' rather than stacked adjacency matrices. By switching the \code{type} argument from \code{"tolist"} into \code{"toarray"} (or \code{"toarray"} for 2-mode data), it is possible to select particular bundle classes, and below is an example to identify the tie entrainment patterns in this particular network.
\begin{Schunk}
\begin{Sinput}
R> rel.sys(netC, type = "toarray", bonds = "tent")
\end{Sinput}
\begin{Soutput}[fontsize=\small]
, , C

    357 398
357   0   0
398   1   0

, , F

    357 398
357   0   0
398   1   0

, , K

    357 398
357   0   0
398   0   0
\end{Soutput}
\end{Schunk}

A system of \emph{strong bonds} of the network is obtained either with a specific command or by selecting the bundle types with a mutual character. These are the reciprocals, tie exchange, and the mixed patterns occurring in the network. Likewise, a system made of \emph{weak bonds} comprise bundle patterns with an asymmetric character that typically characterizes hierarchical structures; that is, asymmetric dyads and tie entrainment bundles. 

However, what establishes the character of the bundle patterns is the relational content, and the categorization of social ties made by T{\"o}nnies \citeyearpar{Tonnies1940} as ``affective'' and ``instrumental'' (or \emph{Gemeinschaft} and \emph{Gesellschaft} in his original formulation) can provide some clues toward a topology of relational bundles. While the entrainment of any category of ties would produce an asymmetric pattern, the exchange of only affective or just instrumental positively valuated ties would yield in a mutual character. Yet the altercation of different categories of ties would not necessarily be a mutual bundle, since the relational content of the ties at the end may suggest a different story.

\begin{Schunk}
\begin{Sinput}
R> identical(
+    rel.sys(netC, type = "toarray", bonds = c("recp", "txch", "mixd")),
+    rel.sys(netC, type = "toarray", bonds = "strong")  )
\end{Sinput}
\begin{Soutput}[fontsize=\small]
[1] TRUE
\end{Soutput}
\end{Schunk}

Bundle patterns are important building blocks for creating theories in multiplex social networks. For instance, the visualization of the strong bonds system corresponding to \code{netC} as an undirected multigraph (not shown here) is possible with the ties of \code{rel.sys} in the code below, and we will notice that all bundles with a mutual character are occurring in the same component of the entire network. It is then possible from such arrangement to make a substantial interpretation of the complex structure, and a social influence process through direct contacts ---for example--- is more likely to happen just within this part of the network.

\begin{Sinput}
R> multigraph(rel.sys(netC, bonds = "strong")$Ties, directed = FALSE)
\end{Sinput}

\subsection{Statistical approach to bundle patterns}
Bundle patterns allow us modeling structural features of multiplex network structures in stochastic terms. For instance, \citet{Wass80} proposed a simple stochastic model for measuring both the level of ``cohesion'' and ``reciprocity'' in a simple network that is based on the three dyadic parameters studied by \citet{Hol-Lei81}.

The maximum likelihood estimate for \emph{group cohesion} is the proportion of asymmetric dyads in the network, $\aA$, to twice the amount of null dyads, $\nN$ \citep[cf. also][]{ProcLoom51}. That is, $$\widehat{\gamma}_{\;\text{cohesion}} = \frac{\aA}{2\cdot\nN},$$ where $\aA = (i,j) \in R \wedge (j,i) \notin R$, and $\nN = (i,j) \notin R \wedge (j,i) \notin R$, for $i>j$ and $r=1$.


The \emph{reciprocity} level in the network is defined by the log odds of the ratio of group ``coherence'', which is the proportion of twice the mutual dyads $\mM$ and asymmetric patterns, to the score for group cohesion. This is the log odds ratio $$\log\Bigl(\frac{\widehat{\gamma}_{\;\text{coherence}}}{\widehat{\gamma}_{\;\text{cohesion}}}\Bigr)$$ where $$\widehat{\gamma}_{\;\text{coherence}} = \frac{2\cdot\mM}{\aA}$$ and $\mM = (i,j) \in R \wedge (j,i) \in R$.


Since weak and strong bonds generalize asymmetric and reciprocal dyads, respectively, the estimation of group cohesion and the reciprocity level in multiplex network structures is straightforward by counting in the bundle census the amount of strong and weak bonds: 
\begin{equation*}\label{eq:Cohesion}
\begin{split}
\text{Cohesion} \;=\; \frac{\#\text{ weak bonds}}{2 \cdot \#\text{ null bundles}}
\end{split}
\end{equation*}
\vspace{0pt}%
\begin{equation*}\label{eq:Reciprocity}
\begin{split}
\text{Reciprocity} \;=\; \log\Biggl(\frac{\frac{2 \;\cdot\; \#\text{ strong bonds}}{\#\text{ weak bonds}}}{\text{Cohesion}}\Biggr),
\end{split}
\end{equation*}
where the amount of null bundles in the proportion for cohesion for a network of $n$ actors equals 
\begin{equation*}\label{eq:Null}
\#\text{ null bundles}  \;=\;  \binom{n}{2} \;-\; \#\text{ strong and weak bonds}.
\end{equation*}


\subsection{Semigroup of relations}
While the wide extension to the link corresponds to the bundle patterns with a multiple character, the long extension is expressed algebraically by the concatenation of the existing ties in the network. The algebraic object ``semigroup'' serves to represent the logic of interlock of network relations, which are the links among simple and concatenated ties.

Formally, the \emph{semigroup of relations} is made of a set of elements $S$ of string relations together with an endowed operation `$\circ$' to the set corresponding to the concatenation of ties:

$$
\bigl\langle\quad S, \quad \circ \quad\bigr\rangle
$$

Each semigroup $S$ is closed under the operation, which means that the product of two or more elements in $S$ must be part of the semigroup. The concatenation of ties is a binary operation on an ordered pair $\circ\colon S \times S \to S$, that for all $x, y, z \in S$ satisfies the Associative Law; i.e., $x \circ (y \circ z) = (x \circ y) \circ z$.

The Zermelo-Fraenkel Axiom of Extensionality, which is paraphrased as the ``Axiom of Quality'' by \citet{BoorWhite1976}, ensures that the semigroup is limited to a number of unique string relations, and this is because it equates the ties that link precisely the same individuals in the network:
$$
R_1 = R_2 \;\;\text{implies}\;\; R_1 \leq R_2 \;\text{and}\; R_2 \leq R_1.
$$
where $\leq$ is a partial order relation that is reflexive, transitive, and antisymmetric. The application of the Axiom of Quality makes possible that just some representative strings characterize $S$ or the network relational structure.
The partial order relation between a pair of representative strings $R_1 \leq R_2$ exists iff relation $R_2$ contains relation $R_1$; that is, for all $(i, j) \in \Xx$, $(i, j) \in R_1 \;\;\text{implies}\;\; (i, j) \in R_2$.

Diverse algebraic constrictions existing in the relational structure of the network is represented by the \emph{partially ordered semigroup} \citep{Pattison1993}, which is defined by the triple:
$$
\bigl\langle\quad S, \quad \circ, \quad \leq \quad\bigr\rangle\text{;}
$$

that is, an \emph{abstract} semigroup with a partial order among its elements where the semigroup is the product of the set of equations implicit in the string classes. On the other hand, the partial order structure or ``poset'' represents a hierarchy among pairwise disjointed subsets of string relations, which is another algebraic restriction. 
Later on, we will look at another kind of algebraic restriction that is based on the existing relations among the categories of strings, and which is produced by means of a decomposition of the semigroup structure.

\subsubsection{Strings and equations}
\pkg{multiplex} offers a suite of functions that allow processing abstract and partially ordered semigroups. To illustrate the semigroup construction, we ``select'' a small component in Incubator network C (cf. Fig.~\ref{fig:incC}) with functions \code{rel.sys} and \code{comps}, and record the small component in object \code{nCc}.

\begin{Sinput}
R> nCc <- rel.sys(netC, type = "toarray", sel = comps(netC)$com[[3]])
\end{Sinput}

Function \code{strings} provides the representative string relations in $S$ ---i.e., primitives and words if they exist--- as a \code{"Strings"} class object. The output is given by taking the first element in a lexicographic order of a set of strings of those linking the same network members. For instance, the \code{st} component provides the labeling of the representative strings, \code{ord} indicates the \emph{order} of $S$, whereas the ``word tables'' with the connections is given as arrays in \code{wt}. 

\begin{Schunk}
\begin{Sinput}
R> strings(nCc)
\end{Sinput}
\begin{Soutput}[fontsize=\small]
$wt
, , C

    339 354 357 395 398
339   0   0   0   0   0
354   0   0   0   0   0
357   0   0   0   0   0
395   0   0   0   0   0
398   0   0   1   0   0
...

$ord
[1] 17

$st
 [1] "C"      "F"      "K"      "CF"     "FC"     "FF"     "CFF"    "FCF"    "FFC"   
[10] "FFF"    "FCFF"   "FFCF"   "FFFC"   "FFFF"   "FFCFF"  "FFFCF"  "FFFCFF"

attr(,"class")
[1] "Strings"
\end{Soutput}
\end{Schunk}

Once we know the unique elements of the semigroup a natural question arise, namely which are the strings that are equated to these representative relations? 
By activating the \code{equat} argument in this function, we apply the Axiom of Quality to the algebraic structure and produce the set of equations with the representative strings of the network relational structure. 
For this, \code{k} serves to establish the length of the compounds involved in the equations, and the attribute names of each vector in the list of component \code{equat} provides the representative string labels. 
In this case, only \code{K} is equated with all compounds until length 3. 

\begin{Schunk}
\begin{Sinput}
R> strings(nCc, equat = TRUE, k = 3)$equat
\end{Sinput}
\begin{Soutput}[fontsize=\small]
$K
 [1] "K"   "CC"  "KK"  "CK"  "KC"  "FK"  "KF"  "CCC" "KKC" "CKK" "CCF" "KKF" "FCC"
[14] "FKK" "CCK" "FFK" "KKK" "KCC" "KFF" "KCK" "CFC" "KFK" "CKC" "FKF" "CFK" "CKF"
[27] "FCK" "FKC" "KCF" "KFC"
\end{Soutput}
\end{Schunk}

\subsubsection{Multiplication table and partial order}
The interrelations among strings is expressed by the semigroup structure, and function \code{semigroup} produces the \emph{multiplication table} of this algebraic structure in the $S$ component of the output. The multiplication table can have either a \code{"numerical"} or a \code{"symbolic"} format, where the former (default) option assigns an integer to each string relation, whereas the symbolic representation assigns a label to each unique element of the semigroup that coincides with the output of the \code{strings} function. Both selections in \code{semigroup} produce a \code{"Semigroup"} class object that is required to perform the decomposition of $S$ afterward.

On the other hand, function \code{partial.order} serves to establish the \emph{partial order} structure of the semigroup provided that the \code{type} argument is set to \code{"strings"}. In such case, the input data of the partial order function is of a \code{"Strings"} class object, and the output is an object \code{"Partial.Order"} of the type chosen class. Both the poset structure and the multiplication table then constitute partially ordered semigroup of the network relational structure.

\begin{Schunk}
\begin{Sinput}
R> semigroup(nCc, type = "numerical")$S
\end{Sinput}
\begin{Soutput}[fontsize=\small]
    1  2 3  4  5  6  7  8  9 10 11 12 13 14 15 16 17
1   3  4 3  3  3  7  3  3  1  1  3  4  3  4  7  3  3
2   5  6 3  8  9 10 11 12 13 14 15 16  5  6 17  8 11
3   3  3 3  3  3  3  3  3  3  3  3  3  3  3  3  3  3
4   3  7 3  3  1  1  3  4  3  4  7  3  3  7  3  3  3
5   3  8 3  3  3 11  3  3  5  5  3  8  3  8 11  3  3
6   9 10 3 12 13 14 15 16  5  6 17  8  9 10 11 12 15
7   1  1 3  4  3  4  7  3  3  7  3  3  1  1  3  4  7
8   3 11 3  3  5  5  3  8  3  8 11  3  3 11  3  3  3
9   3 12 3  3  3 15  3  3  9  9  3 12  3 12 15  3  3
10 13 14 3 16  5  6 17  8  9 10 11 12 13 14 15 16 17
11  5  5 3  8  3  8 11  3  3 11  3  3  5  5  3  8 11
12  3 15 3  3  9  9  3 12  3 12 15  3  3 15  3  3  3
13  3 16 3  3  3 17  3  3 13 13  3 16  3 16 17  3  3
14  5  6 3  8  9 10 11 12 13 14 15 16  5  6 17  8 11
15  9  9 3 12  3 12 15  3  3 15  3  3  9  9  3 12 15
16  3 17 3  3 13 13  3 16  3 16 17  3  3 17  3  3  3
17 13 13 3 16  3 16 17  3  3 17  3  3 13 13  3 16 17
\end{Soutput}
\end{Schunk}

\begin{Schunk}
\begin{Sinput}
R> partial.order(strings(nCc), type="strings")
\end{Sinput}
\begin{Soutput}[fontsize=\small]
       C F K CF FC FF CFF FCF FFC FFF FCFF FFCF FFFC FFFF FFCFF FFFCF FFFCFF
C      1 1 0  0  0  0   0   0   0   0    0    0    1    1     0     0      0
F      0 1 0  0  0  0   0   0   0   0    0    0    0    0     0     0      0
K      1 1 1  1  1  1   1   1   1   1    1    1    1    1     1     1      1
CF     0 0 0  1  0  1   0   0   0   0    0    0    0    0     0     1      0
FC     0 0 0  0  1  1   0   0   0   0    0    0    0    0     0     0      0
FF     0 0 0  0  0  1   0   0   0   0    0    0    0    0     0     0      0
CFF    0 0 0  0  0  0   1   0   0   1    0    0    0    0     0     0      1
FCF    0 0 0  0  0  0   0   1   0   1    0    0    0    0     0     0      0
FFC    0 0 0  0  0  0   0   0   1   1    0    0    0    0     0     0      0
FFF    0 0 0  0  0  0   0   0   0   1    0    0    0    0     0     0      0
FCFF   0 1 0  0  0  0   0   0   0   0    1    0    0    1     0     0      0
FFCF   0 0 0  0  0  0   0   0   0   0    0    1    0    1     0     0      0
FFFC   0 0 0  0  0  0   0   0   0   0    0    0    1    1     0     0      0
FFFF   0 0 0  0  0  0   0   0   0   0    0    0    0    1     0     0      0
FFCFF  0 0 0  0  0  1   0   0   0   0    0    0    0    0     1     0      0
FFFCF  0 0 0  0  0  1   0   0   0   0    0    0    0    0     0     1      0
FFFCFF 0 0 0  0  0  0   0   0   0   1    0    0    0    0     0     0      1
\end{Soutput}
\end{Schunk}

\subsection{Positional analysis and role structure}
Even with a small network, the partially ordered semigroup of the social system is likely to become large and complex. 
Thus, rather than focusing on relational structures based on strings among the individual actors, one way to gain better insight into the relational interlock of a relative large real-world social network is first to construct the network positional system instead for. 
``Structurally equivalent'' actors in the network who are meant to occupy the same ``position'' in the system make such arrangement, and in this way, we are able to reduce dramatically the size of the social system without losing its essential structure, which is a crucial feature for further analyzes and for the substantial interpretation of the network relational interlock. 
Since the actors in an alike position are meant to play a similar role in the social system, the relational structure of a positional system is called the \emph{role structure} of the network. 

A key aspect in network reduction is the definition of structurally equivalence among the actors. There are several meanings of structurally correspondence in social network analysis and it is outside the scope of this paper to look at the different definitions; however, it is important to mention that most of the equivalence notions are designed for simple networks with a ``global'' perspective, and some definitions are stricter criteria than others \citep[see e.g.,][for widespread equivalence types]{LorrWht71, DorBatFer2004}. For multiplex networks, however, we need to take into account the multiplicity of the ties in the reduction of the network, and this can be achieved by means of a ``local'' perspective in the equivalence definition, i.e., the point of view of individual actors.

One correspondence type with a local perspective suitable for multiplex networks is \emph{local role equivalence} \citep{WinMan83}, which is based on the role relations of the individual actors in the network and their respective role sets. This information is stored in a three-dimensional array called \emph{Relation-Box}, which is a device where the generators and compound relations of the network are bounded together. Because social actors are typically unaware of the long chains of relations surrounding them, the researcher can define a ``truncated'' version of the Relation-Box with composite ties until a certain length $k$.

\subsubsection{Compositional equivalence}
\citet{BrePat86} developed a correspondence type based on the local role equivalence since the establishment of roles and positions in the network is made with the perspectives of the individual actors, but considering also the relational features common to all network members. Specific standpoints are partial algebras operationalized in the form of ``person hierarchies'', which are then ``cumulated'' into a single poset structure where the partition of the network takes place. Due the compound relations provide substantial information for the establishment of the positional system, we refer to this correspondence type as \emph{Compositional equivalence}, also known as ``ego algebra'' \citep{WassFau1994}.

Function \code{rbox} produces the Relation-Box of a given network as a \code{"Rel.Box"} class object, which is a three-dimensional array resembling the word tables given by the \code{strings} function but \textit{without} applying the Axiom of Quality to the string relations. This means that none of the primitives nor the compounds are not equated with each other. A hierarchy among the network members is perceived by each actor according to their paths of relations, and this is obtained with the \code{hierar} function provided that the \code{"person"} option selected in the \code{type} argument. For instance, in the case of the first actor of network \code{nCc} the two options below are equivalent for producing this particular person hierarchy.

\begin{Sinput}
R> hierar(rbox(nCc), 1, type = "person")
R> hierar(rbox(nCc), "339", type = "person")
\end{Sinput}

A significant aspect in the definition of compositionally equivalent actors is, however, the accumulation of the network partial algebras, which are represented in the horizontal slices in the Relation-Box, and which are called ``relation planes''. The rows in the relational planes record the collection of string relations for a distinguished individual actor, whereas the columns represent the actors' ``role relations'' that correspond to the ties of a particular kind to the rest of the actors in the system. The aggregation of person hierarchies is expressed by a poset structure known as the \emph{cumulated person hierarchy} or CPH from the set of inclusion relations among relational planes after applying the transitive closure. Since CPH is a poset then the structure is reflexive, transitive, and antisymmetric.

Function \code{cph} serves to construct the cumulated person hierarchy, and relies on a \code{"Rel.Box"} object class as well. The following example shows the aggregation of the partial algebras in \code{nCc} based on the default length of ties in \code{rbox}, and the output for $k=3$ shows that only two of the actors are \emph{comparable}.
\begin{Schunk}
\begin{Sinput}
R> cph(rbox(nCc))
\end{Sinput}
\begin{Soutput}[fontsize=\small]
    339 354 357 395 398
339   1   0   0   0   1
354   0   1   0   0   0
357   0   0   1   0   0
395   0   0   0   1   0
398   0   0   0   0   1
attr(,"class")
[1] "Partial.Order" "CPH"
\end{Soutput}
\end{Schunk}

The construction of the role structure as a partially ordered semigroup relies on the CPH with the chosen length of strings, and now we review the whole process of establishing the network positional system and role structure of an empirical network represented by Incubator C, \code{netC}. In this case the ties of the network are directed, and for digraphs is suggested to generate first ``relational contrast'' in the system, which is achieved by including the ``tie transposes'' in the construction of the Relation-Box. Thus, if relation $\cC$ in the network represents ``collaborates with,'' then the tie transpose will stand for ``pointed as collaborator by.'' A similar construction occurs with the other kinds of tie occurring in the system.

To include tie transposes of the primitives in the construction of the Relation-Box, we activate the \code{transp} argument in the \code{rbox} function as with \code{netCrb}. This means that the CPH of this network in \code{netCcph}, which is based on the Relation-Box, reflects tie transposition as well.

\begin{Sinput}
R> netCrb <- rbox(netC, transp = TRUE)
R> netCcph <- cph(netCrb)
\end{Sinput}

With Compositional equivalence, structurally correspondent actors will have a similar set ---or rather lack--- of inclusions in the partial order structure, and visualizing the poset often provides useful insight for the partition of the network, especially when the structure is relatively large. Hence, with the convenient function \code{diagram} that uses the functionalities provided by the \pkg{Rgraphviz} package \citep{Rgraphviz:2018} it is possible to plot the poset representing CPH as an inclusion lattice or ``Hasse'' diagram \citep{DavPri2002}.

\begin{figure*}[t]
\vspace{-15pt}
\begin{tabular}{cc}
    \includegraphics[width=7.5cm,page=2]{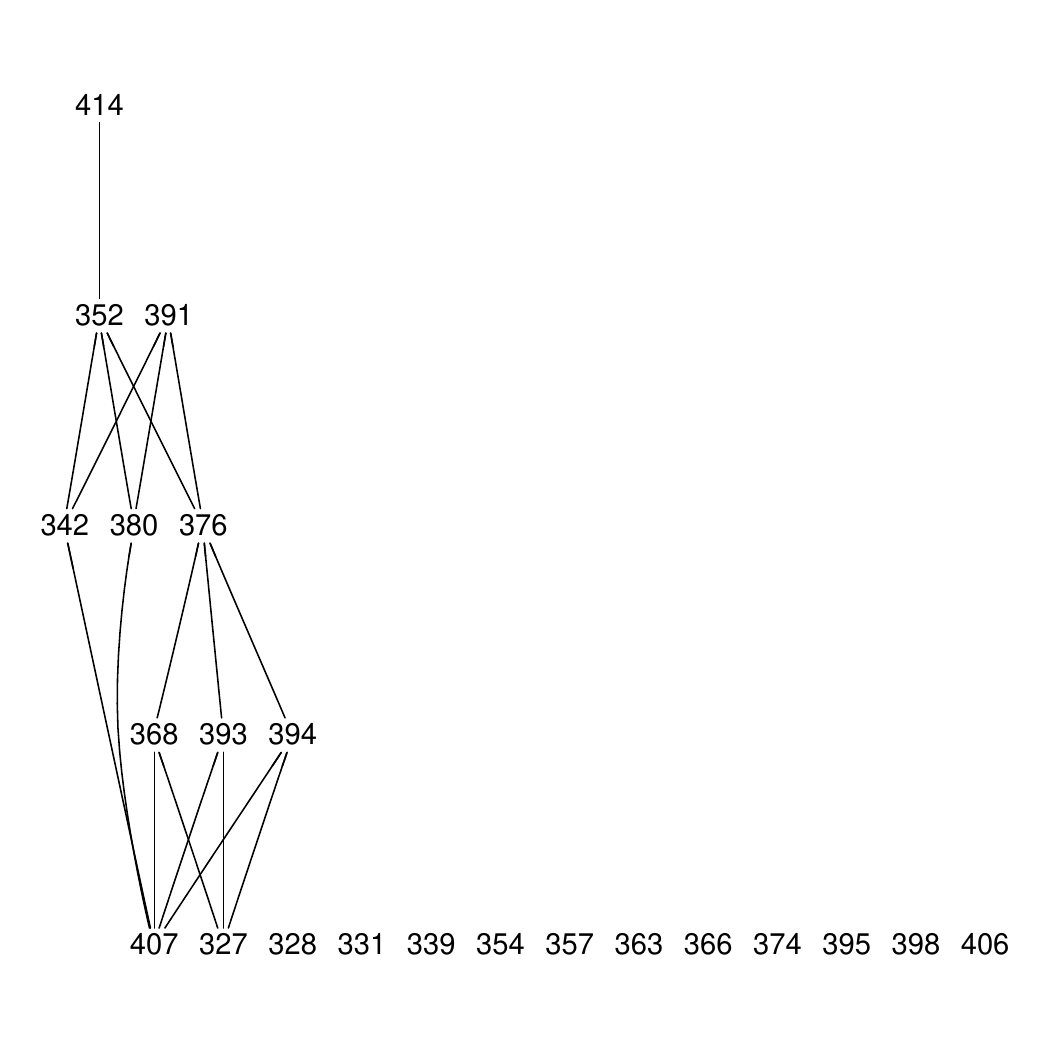}  \; & \;
    \includegraphics[width=7.5cm,page=3]{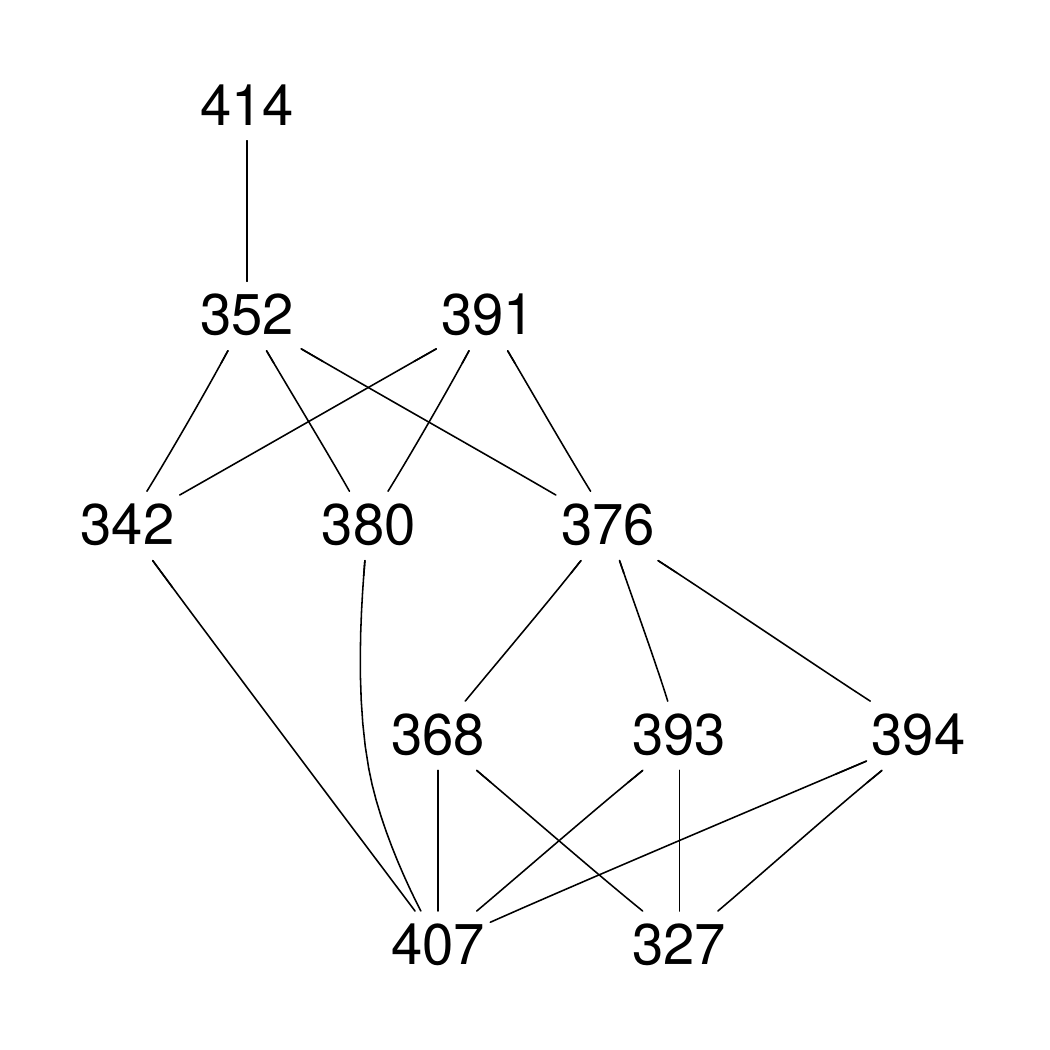}
\end{tabular}
\setlength{\abovecaptionskip}{0pt}
  \caption{CPH of Incubator C with and without incomparable elements in the poset.}
\label{fig:CPHsC}
\end{figure*}

Figure~\ref{fig:CPHsC} shows the inclusion lattice of the CPH of Incubator C plotted with the two flavors available with the \code{diagram} function. The picture to the left includes all elements in the partial order structure, whereas the plot to the right is given without the ``incomparable'' elements in the poset, which is possible by disabling the \code{incmp} argument in the function.

\begin{Sinput}
R> library("Rgraphviz")
R> diagram(netCcph)
R> diagram(netCcph, incmp = FALSE)
\end{Sinput}

As with many cases, there is no univocal solution in the network partition, and the researcher needs to make a judgment of the correspondence between actors that is based on theory or some other criteria. The simplest way of partitioning this particular network would be making two classes of actors, one with those who have an inclusion relation in the CPH, and another class with the incomparable actors. However, most of the times this is a trivial solution since it ends up with the identity and universal matrices, which either have none or an annihilating structuring effect in the role structure, and we need to differentiate equivalent actors who are linked in the inclusion lattice structure as well.

Based on the output from Fig.~\ref{fig:CPHsC} we categorize the actors into three classes---one for the incomparable actors in the poset---and record this information as a vector in object \code{cls}.

\begin{Schunk}
\begin{Sinput}
R> cls <- c(2,3,3,3,2,1,3,3,3,3,2,3,2,2,1,2,2,3,3,3,2,1)
\end{Sinput}
\end{Schunk}
where the class membership can be observed as:
\begin{Schunk}
\begin{Sinput}
R> as.table(rbind(dimnames(netCcph)[[1]], cls))
\end{Sinput}
\begin{Soutput}[fontsize=\footnotesize]
    A   B   C   D   E   F   G   H   I   J   K   L   M   N   O   P   Q   R   S   T   U   V  
    327 328 331 339 342 352 354 357 363 366 368 374 376 380 391 393 394 395 398 406 407 414
cls 2   3   3   3   2   1   3   3   3   3   2   3   2   2   1   2   2   3   3   3   2   1  
\end{Soutput}
\end{Schunk}

Function \code{perm} allows making a permutation of the matrix representing the cumulated person hierarchy with the clustering information allocated to the \code{clu} argument and we verify that actors \textsf{352}, \textsf{391}, and actor \textsf{414} (who covers the first one) clearly differentiate from the rest of the network members. Such differentiation is because these three actors contain the rest of the elements in the poset structure without being contained by them, and the output below involving the linked actors in the lattice structure serves as the basis for the network partition.

\begin{Schunk}
\begin{Sinput}
R> perm(netCcph, clu = cls)
\end{Sinput}
\begin{Soutput}[fontsize=\footnotesize]
    352 391 414 327 342 368 376 380 393 394 407 328 331 339 354 357 363 366 374 395 398 406
352   1   0   1   0   0   0   0   0   0   0   0   0   0   0   0   0   0   0   0   0   0   0
391   0   1   0   0   0   0   0   0   0   0   0   0   0   0   0   0   0   0   0   0   0   0
414   0   0   1   0   0   0   0   0   0   0   0   0   0   0   0   0   0   0   0   0   0   0
327   1   1   1   1   0   1   1   0   1   1   0   0   0   0   0   0   0   0   0   0   0   0
342   1   1   1   0   1   0   0   1   0   0   0   0   0   0   0   0   0   0   0   0   0   0
368   1   1   1   0   0   1   1   0   1   1   0   0   0   0   0   0   0   0   0   0   0   0
376   1   1   1   0   0   0   1   0   0   0   0   0   0   0   0   0   0   0   0   0   0   0
380   1   1   1   0   1   0   0   1   0   0   0   0   0   0   0   0   0   0   0   0   0   0
393   1   1   1   0   0   1   1   0   1   1   0   0   0   0   0   0   0   0   0   0   0   0
394   1   1   1   0   0   1   1   0   1   1   0   0   0   0   0   0   0   0   0   0   0   0
407   1   1   1   0   1   1   1   1   1   1   1   0   0   0   0   0   0   0   0   0   0   0
...
\end{Soutput}
\end{Schunk}

\subsection{Algebraic constraints}
Although it is possible to perform further partitions of the poset leading to a larger structure, a single- actor position seldom has a role in social life, and collective actors typically play social roles. Moreover, a large structure will usually ---but not necessarily--- produce a large semigroup of role relations, which is something we try to avoid in the first place. As a result, the clustering information used for the permutation of the CPH is applied in the reduction of the network by function \code{reduc}, which establishes the \emph{positional system} of the network recorded in object \code{netCps}. 

With directed multiplex networks, there are three kinds of \emph{algebraic constraints}: One constitutes the ``role table'' representing the semigroup where the collective ties are interrelated. This structure is obtained with the \code{semigroup} function as well, but now with the network positional system as the generator relations \code{gens} in the semigroup.

\begin{Schunk}
\begin{Sinput}
R> netCps <- reduc(netC, clu = cls)
R> netCS <- semigroup(netCps, type = "symbolic")
\end{Sinput}
\begin{Soutput}[fontsize=\small]
...
$gens
, , C

  2 3 1
2 1 1 0
3 1 1 1
1 0 1 1

, , F

  2 3 1
2 1 1 0
3 1 1 0
1 0 1 1

, , K

  2 3 1
2 0 0 0
3 0 1 0
1 0 1 0

...

$S
     C   F  K CC CF CK  FF KC  KF CKF
C   CC  CF CK CC CC CK  CF CC CKF CKF
F   CC  FF CK CC CC CK  FF CC CKF CKF
K   KC  KF  K KC KC  K  KF KC  KF  KF
CC  CC  CC CK CC CC CK  CC CC CKF CKF
CF  CC  CF CK CC CC CK  CF CC CKF CKF
CK  CC CKF CK CC CC CK CKF CC CKF CKF
FF  CC  FF CK CC CC CK  FF CC CKF CKF
KC  KC  KC  K KC KC  K  KC KC  KF  KF
KF  KC  KF  K KC KC  K  KF KC  KF  KF
CKF CC CKF CK CC CC CK CKF CC CKF CKF

attr(,"class")
[1] "Semigroup" "symbolic"
\end{Soutput}
\end{Schunk}

\begin{figure*}[th!]
\vspace{-35pt}
    \centering\includegraphics[width=10.5cm,page=4]{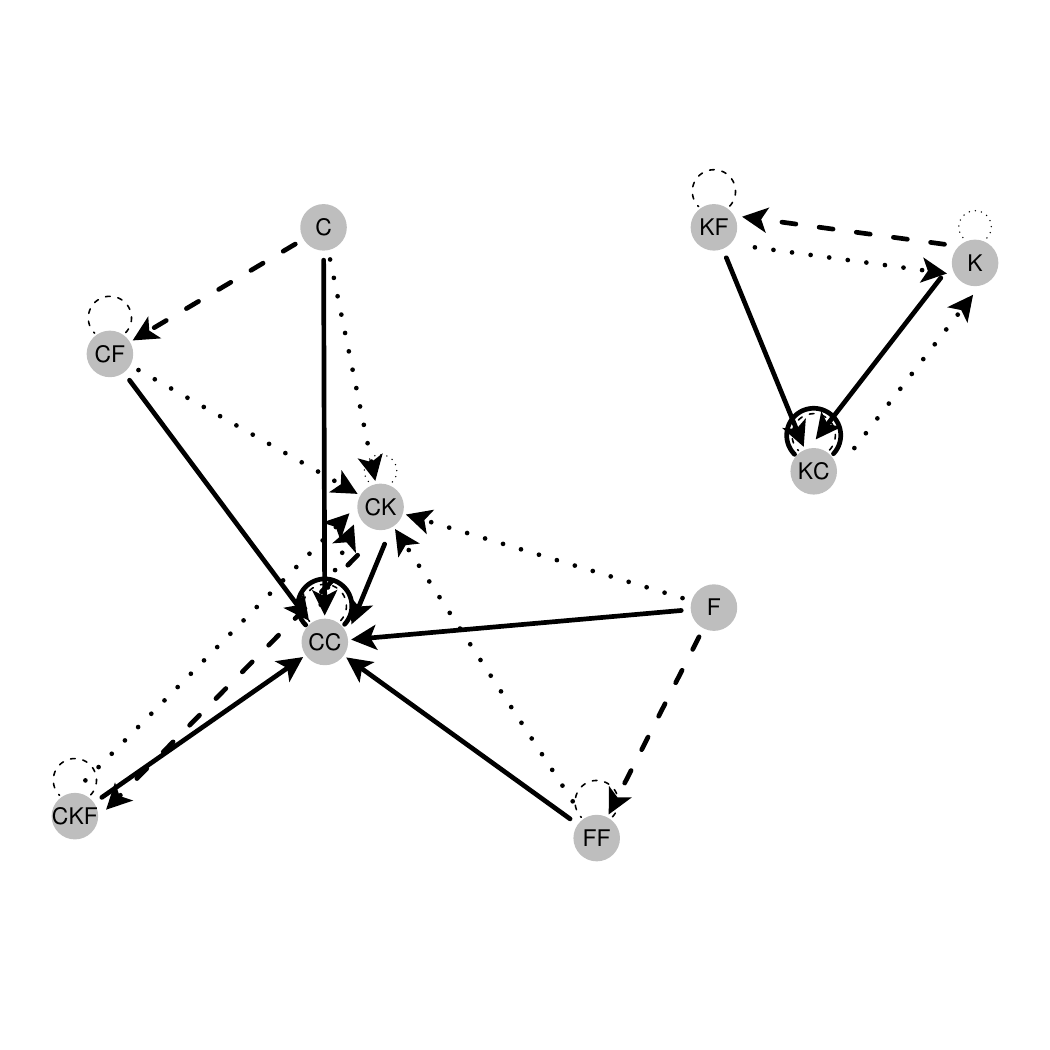}
\setlength{\abovecaptionskip}{-25pt}
  \caption{Cayley colour graph of the semigroup structure in \code{netCS}.}
\label{fig:CCgraph}
\end{figure*}

The \emph{Cayley colour graph} or just Cayley graph is a graphical representation of the relationships among the relations in the network relational structure, which is given in this case as a semigroup of role relations with a symbolic format. Figure~\ref{fig:CCgraph} depicts the Cayley graph of \code{netCS}, which is produced with the \pkg{multigraph} function \code{ccgraph} and the code below. Each generator is represented by a particular arc shape, and each representative string is represented by a node. We notice that there are two components, one for relation \code{K}, and the other for \code{C} and \code{F}; both with compounds product of ``right multiplication''.

\begin{Sinput}
R> scpc <- list(cex = 4, vcol = 8, lwd = 3, ecol = 1, fsize = 9, pos = 0)
R> ccgraph(netCS, seed = 1, scope = scpc)
\end{Sinput}

Even though the representative strings in the role structure is given in component \code{st} of the output of the semigroup routine, once again we use function \code{strings} in the context of the network positional system to obtain the ``set of equations'', which is another algebraic constraint. We obtain the equations of strings until length 3 with the code below, and we observe for example in the output that one generator ---namely \code{K}--- is an idempotent element in the semigroup. In general, elements having the idempotence property are significant in finding subgroups in this algebraic system \citep{Boyd1991}, which are more regular structures. The other equations occur just among compounds relations where \code{FF} is an idempotent element. 

\begin{Schunk}
\begin{Sinput}
R> strings(netCps, equat = TRUE, k = 3)$equat
\end{Sinput}
\begin{Soutput}[fontsize=\small]
$equat
$equat$K
[1] "K"   "KK"  "KKK" "KCK" "KFK"

$equat$CC
 [1] "CC"  "FC"  "CCC" "FFC" "CCF" "FCC" "FCF" "CFC" "CKC" "FKC"

$equat$CF
[1] "CF"  "CFF"

$equat$CK
[1] "CK"  "FK"  "CKK" "FKK" "CCK" "FFK" "CFK" "FCK"

$equat$FF
[1] "FF"  "FFF"

$equat$KC
[1] "KC"  "KKC" "KCC" "KCF" "KFC"

$equat$KF
[1] "KF"  "KKF" "KFF"

$equat$CKF
[1] "CKF" "FKF"
\end{Soutput}
\end{Schunk}

\begin{figure*}[t]
    \centering\includegraphics[width=9.2cm,page=5]{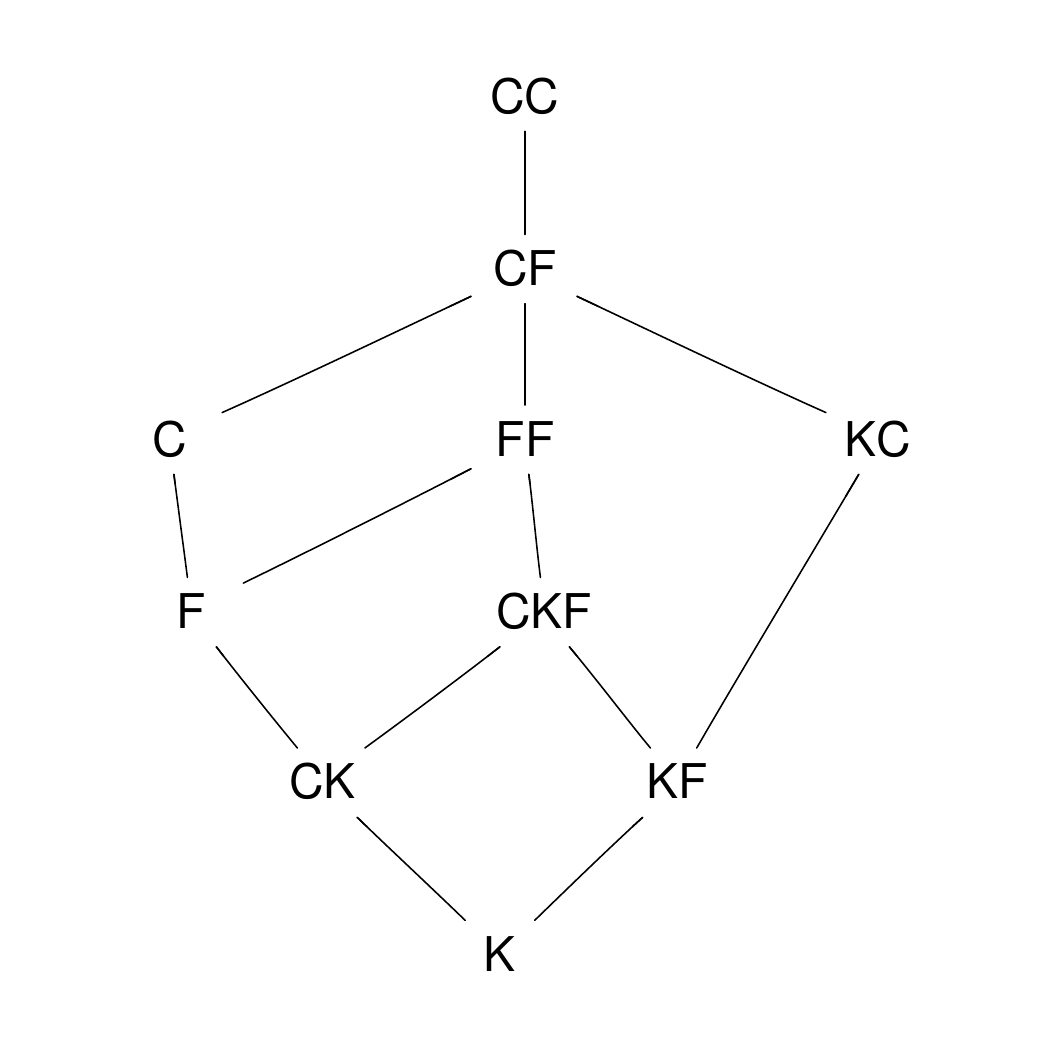}
\setlength{\abovecaptionskip}{10pt}
  \caption{Hierarchy of string relations in the role structure of \code{netC}.}
\label{fig:PosetRSnetC}
\end{figure*}

As seen before, the set of equations provides significant information when it comes to interpreting the hierarchy of the string relations, and this is because the poset structure is based on the unique strings computed with the \code{strings} function, as the objects \code{netCst} and \code{netCpo} reflect it. The depiction of partial order of the role structure is a convenient way to analyze the third algebraic constraint, namely the ``hierarchy of role relations'' of the network, which for Incubator C is depicted in Figure~\ref{fig:PosetRSnetC} as an inclusion lattice with the \code{diagram} function. In the inclusion lattice or Hasse diagram, the supremum role relations cover infimum elements labeled with the representative strings, and we notice in the poset of the role structure that the lattice structure has a compound and a primitive role as the maximal and the minimal elements, respectively.

\begin{Sinput}
R> netCpo <- partial.order(strings(netCps), type = "strings")
R> diagram(netCpo)
\end{Sinput}

A substantial interpretation of these algebraic constraints states that if $\fF$ represents an informal friendship role relation, and $\kK$ perceived competition then are given within a formal collaboration network; that is, $\fF \leq \cC$ and $\kK \leq \cC$. These role relations occur inside a structure where the friends' collaborators network, $\fF\cC$ (and collaborators' collaborators, $\cC\cC$) that encompasses the arrangement of collaborators' friends, $\cC\fF$. 

However, since the structure is not a lineal order the ``friends of my friends'' ties, $\fF\fF$ occur within the formal-informal pattern ``friends of my collaborators,'' $\cC\fF$ (and ``friends' friends of my collaborators,'' $\cC\fF\fF$), but not necessarily within the ``collaborators of my competitors,'' $\kK\cC$ and the other compounds equated to this pattern. A similar reading is applied to the rest of the strings in the hierarchy of role structure and the analysis of such role interlock for this particular network is made in a systematic way.

It is important to mention that since we included the transposition of the primitive ties, all these relations should be reflected in this arrangement as well. However, despite tie transpositions are not involved in the poset diagram just depicted, each of the inclusions involving $\cC$, $\fF$, and $\kK$ is respected in the hierarchy of the role relations with transposes. 

The role interlock of Incubator C has a set of univocal containments enclosing the generators and the equations of the shortest compounds, where the remaining four strings in the role structure are contained just in the two upper elements where the pair $\cC\fF\kK \leq \fF\fF$ contains $\cC\kK$, and the pair $\kK\fF \leq \kK\cC$ contains just the minimal element.
$$
\kK \;\leq\; ( \cC\kK = \fF\kK ) \;\leq\; \fF \;\leq\; \cC \;\leq\; \cC\fF \;\leq\; ( \cC\cC = \fF\cC ).
$$

\subsection{Decomposition}
The last step in the analysis is to perform a \emph{decomposition} of the semigroup of role relations, most of times by means of a ``subdirect'' representation. The decomposition process produces an ``aggregated'' role structure where string relations are clustered whenever they conform the rules of the ``substitution property'' \citep{HartStea1966}.

In the case of abstract semigroups, functions \code{decomp} and \code{cngr} perform the decomposition of semigroup structures, and below we generate the clustering of strings by means of congruence relations.

\begin{Schunk}
\begin{Sinput}
R> decomp(netCS, cngr(netCS, uniq = TRUE), type = "cc")$clu
\end{Sinput}
\begin{Soutput}[fontsize=\small]
[[1]]
  C   F   K  CC  CF  CK  FF  KC  KF CKF 
  1   3   4   1   1   5   3   6   7   2 

[[2]]
  C   F   K  CC  CF  CK  FF  KC  KF CKF 
  1   1   2   1   1   3   1   4   4   1 

[[3]]
  C   F   K  CC  CF  CK  FF  KC  KF CKF 
  1   3   4   1   1   4   3   1   2   2 

[[4]]
  C   F   K  CC  CF  CK  FF  KC  KF CKF 
  1   1   2   1   1   2   1   1   1   1
\end{Soutput}
\end{Schunk}

For partially ordered semigroups, the decomposition lies on the ``factorization'' of the semigroup structure \citep{Pattison1993}, and this process has two parts. First, function \code{fact} produces the induced inclusions to the partial order, the atoms of the \emph{congruence lattice}, and also the meet-complements of these atoms, whereas the second part of the decomposition is performed with the routine \code{pi.rels} that constructs the partition or $\pi$-relations for $S$. 

Function \code{decomp} below activates the \code{"mca"} \code{type} option that stands for the ``meet-complements of the atoms'', and which are $\pi$-relations closer to the 1-element semigroup, which is the \emph{suprema} or maximal element in the congruence lattice. It is also possible to perform the decomposition of $S$ with other $\pi$-relations in the congruence lattice included the atoms, but the partitions closer to partial order, which is the \emph{infima} or minimal element in the congruence lattice are ``finer'' than the $\pi$-relations closer to the suprema like the meet-complements.

\begin{Schunk}
\begin{Sinput}
R> decomp(netCS, pi.rels(fact(netCS, netCpo)), type = "mca")$clu
\end{Sinput}
\begin{Soutput}[fontsize=\small]
[[1]]
  C   F   K  CC  CF  CK  FF  KC  KF CKF 
  1   2   3   4   5   6   7   8   3   6 

[[2]]
  C   F   K  CC  CF  CK  FF  KC  KF CKF 
  1   2   3   4   4   3   4   4   4   4
\end{Soutput}
\end{Schunk}

Since both representations constitute a subdirect representation of the network role structure, we bear in mind again that these structures are partly overlapped to each other, even though we are confident at the same time that the output is a true representation of the relational structure.


\section[Signed networks]{Signed networks}
A \emph{signed network} is a special type of multiplex structure having particular types of relations among the actors with a different \emph{sign} or \emph{valence}. Signs of ties in signed networks are typically either ``positive'' or ``negative'' such as like and dislike, or collaboration and competition ties between companies or even nations; however, although the prototypical signed structures only have these two contrasting valences, real life social networks can have relationships in which both signs occur simultaneously. Hence, besides positive and negative ties, an ``ambivalent'' relation constitutes another kind of sign, whereas the ``absence'' of a tie between two actors is regarded as a valence as well.

There is a pair of functions in \pkg{multiplex} designed to represent systems with different valences. For example, the first and second matrices of object \code{nCc} from the previous section are by default considered as positive and negative ties in a given system by the \code{signed} function that produces a \code{"Signed"} class object. Because there is an entrainment of ties in this configuration, letters \code{n} \code{a} and \code{o} represent negative, ambivalent, and absent types of relations, respectively.
\begin{Schunk}
\begin{Sinput}
R> signed(nCc)
\end{Sinput}
\begin{Soutput}[fontsize=\small]
val
[1] o n a

$s
    339 354 357 395 398
339 o   o   o   o   n  
354 o   o   o   n   o  
357 n   o   o   o   o  
395 n   o   o   o   n  
398 o   o   a   o   o  

attr(,"class")
[1] "Signed"
\end{Soutput}
\end{Schunk}

In case that the second array of \code{nCc} represents \code{p}, then we explicit the positive ties in the input as below, and when the system has any ambivalent relation, then a positive and a negative integer and zero will represent the valence types in the signed network.
\begin{Schunk}
\begin{Sinput}
R> signed(nCc[ , , 2], nCc[ , , 1])
\end{Sinput}
\begin{Soutput}[fontsize=\small]
$val
[1] p o a

$s
    339 354 357 395 398
339 o   o   o   o   p  
354 o   o   o   p   o  
357 p   o   o   o   o  
395 p   o   o   o   p  
398 o   o   a   o   o  
...
\end{Soutput}
\end{Schunk}

\subsection{Structural balance and Semirings}
Classical procedures designed for the analysis of signed networks try to find equilibrium ---or lack of it-- in network structures that are the product of the combination of ties with different signs. For instance, \citet{Simmel1950} noticed that a conflict can be a mechanism of integration among social actors, and \citeauthor{Heider2013} \citeyearpar[(1st edn. 1958]{Heider2013} developed further on this idea with the \emph{Structural Balance theory}, which sustains that imbalanced structures have an inherent tension and are prone to change, whereas balanced networks are more steady over time.

Roughly speaking, a strictly balanced system has two mutually exclusive groups of elements in which all within-ties are positive and all between-ties are negative. These rules of polarization, known as the \emph{Structure Theorem} \citep{Cartwright56}, can be extended to more than two groups where a structurally balanced system has clusters of elements with positive within-ties and negative between-relations \citep[cf.][]{Davis67}. This means that a signed network ---besides being balanced or imbalanced--- can be clusterable as well, which is a sort of structurally balanced structure as long these conditions apply.

The operationalization of the Structural Balance theory in signed networks is made with an algebraic approach, and in this case the rules of another algebraic system known as \emph{semiring} serve to evaluate the network in terms of structural balance. A semiring is the combination of an abstract semigroup with identity under multiplication and a commutative monoid under addition, and in formal terms this structure is a quintuple:
$$
\bigl\langle\quad Q, \quad +, \quad \cdot\;, \quad 0, \quad 1 \quad\bigr\rangle
$$
where $Q$ is a non-empty set associated to the addition `$+$' and multiplication `$\cdot$' operations together with a pair of special elements, $0$ and $1$, which are the neutral elements under addition and multiplication, respectively, and $0$ acts as an absorbing element under multiplication as well. Besides, multiplication distributes over addition, which means that for $x, y, z \in Q$; $x \cdot (y + z) = (x \cdot y) + (x \cdot z)$ and $(x + y) \cdot z = (x \cdot z) + (y \cdot z)$.

Semirings are then going to be used for assessing whether a signed network is structurally balanced or not, and the evaluation is performed according to the two operations involved in this type of structure \citep[cf.][]{HarNorCat1965}. Basically, a chain of relations ---whether directed or not-- with an even number of positive edges is considered imbalanced; otherwise the chain is regarded as structurally balanced.

The \code{semiring} function of \pkg{multiplex} assesses signed structures in terms of either a \code{"balance"} or a \code{"cluster"} \code{type} semiring structure \citep*[See][for details]{DorBatFer2004}. With the \pkg{base} function \code{formals} we pay special attention to a number of the arguments of this function for a better understanding how it proceeds:
\begin{Schunk}
\begin{Sinput}
R> formals(semiring)
\end{Sinput}
\begin{Soutput}[fontsize=\small]
...
$type
c("balance", "cluster")

$symclos
[1] TRUE
...
$k
[1] 2
\end{Soutput}
\end{Schunk}

Thus, once specifying one of the two \code{type} rules of the semiring structure, the argument \code{symclos}, which stands for ``symmetric closure'', serves to specify whether the evaluation is made in terms of paths or else by the default option that is chains or semipaths. Argument \code{k} states the length of these compounds, and controlling this interval is a decisive aspect in the assessment of the network in terms of the structural balance theory either with or without tie direction.

\subsection{Checking for balance}
We illustrate now the assessment process in terms of structural balance with the Incubator network C that is depicted in Fig.~\ref{fig:incC}; however, since this network is made up of three types of relations we need to establish first the two contrasting relations. Collaboration ties and perceived competition among the actors seems to have opposite (or at least different) signs, and they will represent the positive and negative valences in the signed structure. This means that we disregard friendship relations for this part of the analysis, and the signed structure of Incubator network C is made of the $\cC$ and $\kK$ ties representing the positive and negative valences in the system. As a result, the first and third arrays representing the two kinds of ties in \code{netC} are then recorded in object \code{netC2}

\begin{Sinput}
R> netC2 <- netC[ , , c(1, 3)]
\end{Sinput}

\begin{figure*}[t]
\vspace{-130pt}
    \centering\includegraphics[width=15cm,page=6]{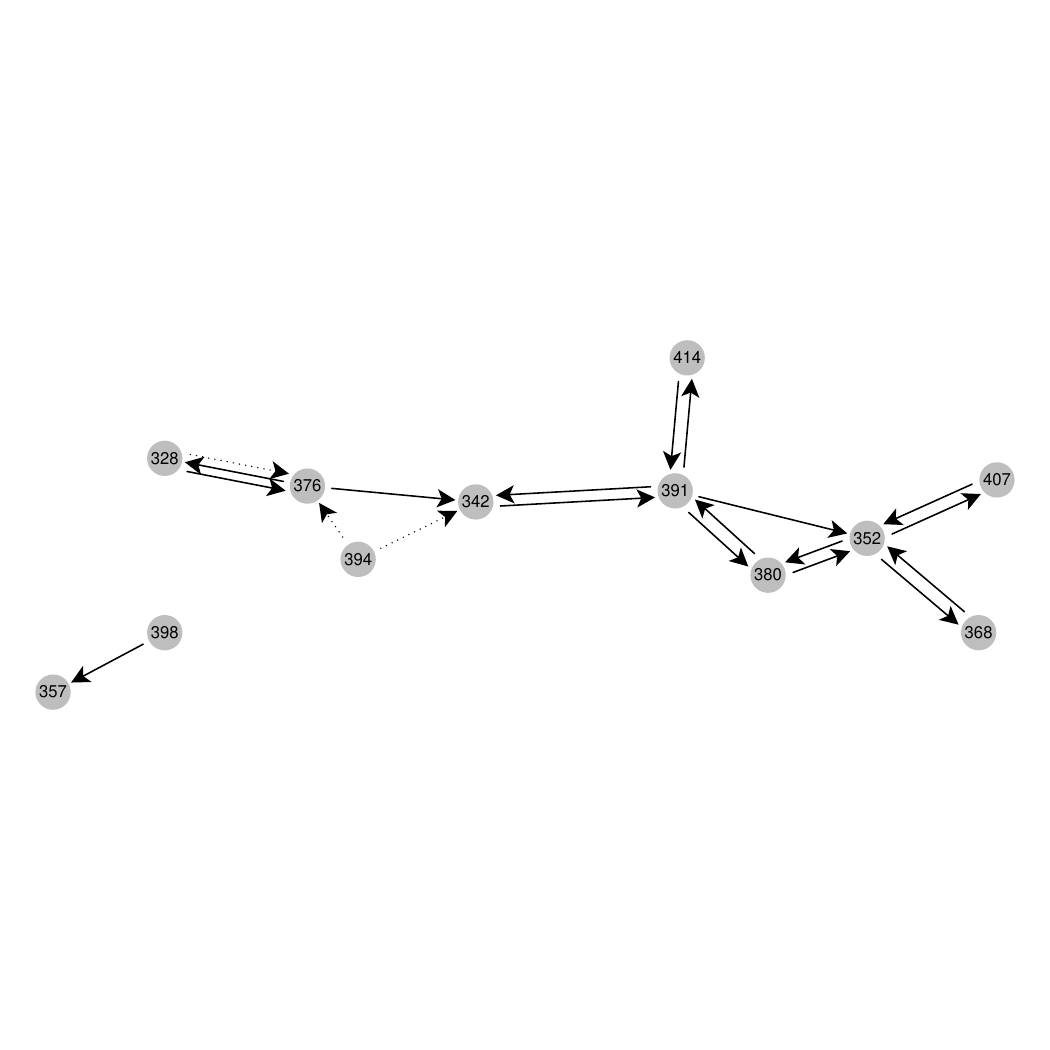}
\setlength{\abovecaptionskip}{-115pt}
  \caption{Multigraph of the signed structure from Incubator C without isolates.}
\label{fig:netCsg}
\end{figure*}

The signed network of \code{netC2} is depicted as a multigraph in Figure~\ref{fig:netCsg} with the \code{multigraph} function and activating the \code{signed} argument to depict the positive and negative ties as solid and dotted arcs, respectively. In this case, we remove the isolated actors from these relation types, which is achieved by applying the \code{rm.isol} function to the network object. 

\begin{Sinput}
R> multigraph(rm.isol(netC2), signed = TRUE, layout = "force", seed = 1, 
+    scope = scp)
\end{Sinput}

First we notice that the system is disconnected and, since is not possible to have chains of relations between actors in different components of the network plus the dyadic component in the multigraph has a single positive tie, the analysis of structural balance focuses on the large component of this network where the two types of tie occur. Function \code{comps} allows listing the members in each of the components in the signed or any network plus the isolated actors, and such output is useful for the extraction of particular actors from the entire system.

\begin{Schunk}
\begin{Sinput}
R> comps(netC2)
\end{Sinput}
\begin{Soutput}[fontsize=\small]
$com
$com[[1]]
 [1] "368" "376" "380" "391" "394" "328" "407" "414" "342" "352"

$com[[2]]
[1] "398" "357"


$isol
 [1] "327" "331" "339" "354" "363" "366" "374" "393" "395" "406"
\end{Soutput}
\end{Schunk}

Because the actors in \code{netC2} bear labels, the construction of the signed structure is made by locating the members inside the first component of the network, and assigning such information into the vector \code{sel} below. The creation of the signed structure is made by applying the \code{signed} function to this particular array of selected elements that represents the large component of the network. The signed network is then recorded as a \code{"Signed"} class object in \code{netCsg} that accounts just for the involved actors in the large component with the collaboration and competition ties.

\begin{Schunk}
\begin{Sinput}
R> sel <- which(dimnames(netC2)[[1]] %in% comps(netC2)$com[[1]])
R> netCsg <- signed(netC2[sel, sel, ])
\end{Sinput}
\begin{Soutput}[fontsize=\small]
$val
[1] p o n a

$s
    328 342 352 368 376 380 391 394 407 414
328 o   o   o   o   a   o   o   o   o   o  
342 o   o   o   o   o   o   p   o   o   o  
352 o   o   o   p   o   p   o   o   p   o  
368 o   o   p   o   o   o   o   o   o   o  
376 p   p   o   o   o   o   o   o   o   o  
380 o   o   p   o   o   o   p   o   o   o  
391 o   p   p   o   o   p   o   o   o   p  
394 o   n   o   o   n   o   o   o   o   o  
407 o   o   p   o   o   o   o   o   o   o  
414 o   o   o   o   o   o   p   o   o   o  

attr(,"class")
[1] "Signed"
\end{Soutput}
\end{Schunk}

We can obtain the signed structure with semipaths in the \code{Q} component of the \code{semiring} function by activating the \code{symclos} argument and setting \code{k} to one.

\begin{Schunk}
\begin{Sinput}
R> semiring(netCsg, symclos = TRUE, k = 1)$Q
\end{Sinput}
\begin{Soutput}[fontsize=\small]
    328 342 352 368 376 380 391 394 407 414
328 o   o   o   o   p   o   o   o   o   o  
342 o   o   o   o   p   o   p   n   o   o  
352 o   o   o   p   o   p   p   o   p   o  
368 o   o   p   o   o   o   o   o   o   o  
376 p   p   o   o   o   o   o   n   o   o  
380 o   o   p   o   o   o   p   o   o   o  
391 o   p   p   o   o   p   o   o   o   p  
394 o   n   o   o   n   o   o   o   o   o  
407 o   o   p   o   o   o   o   o   o   o  
414 o   o   o   o   o   o   p   o   o   o
\end{Soutput}
\end{Schunk}

As we can see in the first signed matrix above, there exists an ambivalent relation from actor \textsf{328} to actor \textsf{376}, which means that the system cannot be strictly balanced when considering paths; i.e., it is not possible to have groups of actors with positive and negative ties only. Nevertheless, in the case of semipaths (second signed matrix) the symmetric closure of the network implies that such bundle pattern involving the ambivalent tie ends up having a positive sign, and this is due to the fact that with this operation a positive (and negative) valence prevails over the ambivalence, and any valence type has prevalence over absent relations.

\subsubsection{Balance semiring}
The analysis of the signed network represented by \code{netCsg} starts with the default options of the \code{semiring} function. This means that the evaluation of the signed structure is made with the rules of a \code{"balance"} \code{type} semiring object and 2-chain relations.

\begin{Schunk}
\begin{Sinput}
R> semiring(netCsg, type = "balance")$Q
\end{Sinput}
\begin{Soutput}[fontsize=\small]
    328 342 352 368 376 380 391 394 407 414
328 p   p   p   p   p   p   p   n   p   p  
342 p   p   p   p   p   p   p   n   p   p  
352 p   p   p   p   p   p   p   o   p   p  
368 p   p   p   p   p   p   p   o   p   p  
376 p   p   p   p   p   p   p   n   p   p  
380 p   p   p   p   p   p   p   o   p   p  
391 p   p   p   p   p   p   p   n   p   p  
394 n   n   o   o   n   o   n   p   o   o  
407 p   p   p   p   p   p   p   o   p   p  
414 p   p   p   p   p   p   p   o   p   p  
\end{Soutput}
\end{Schunk}

Important information is given in the diagonal of the matrices that represents the closed chains of relations in the system, and this is because a structurally balance system will have this vector fully populated with positive entries. In the above case, all values in the diagonal are positive, and the resulted structure suggests that the system has two classes of actors with one class having a single member; such arrangement is viable even if absent chains replace the negative link that should exist in a strictly balanced structure. Since the ambivalent relation involving actors \textsf{328} and \textsf{376} disappears with the symmetric closure, none of the chains involving these actors results ambivalent with the semipath option.

Equally, the examination of the signed structure with paths instead of semipaths is produced by disabling the \code{symclos} argument in the \code{semiring} function as below. In this case, the result is a different arrangement mostly for two . Firstly, because the 2-paths involving \code{a} are ambivalent due the absorbing character of this valence type, and second because there are no incoming ties toward actor \textsf{394}. This means that is not possible to have cycles, i.e., closed paths, involving this particular node but just semicycles or closed semipaths. Moreover, the cycle concerning actor \textsf{328} results ambivalent because the tie originated by this actor is ambivalent, whereas the negative character of the 2-paths involving actor \textsf{394} is due the fact  there is an even number of negative ties in the chain relations.

\begin{Schunk}
\begin{Sinput}
R> semiring(netCsg, type = "balance", symclos = FALSE)$Q
\end{Sinput}
\begin{Soutput}[fontsize=\small]
    328 342 352 368 376 380 391 394 407 414
328 a   a   o   o   o   o   o   o   o   o  
342 o   p   p   p   o   p   p   o   p   p  
352 o   p   p   p   o   p   p   o   p   p  
368 o   p   p   p   o   p   p   o   p   p  
376 o   p   p   p   a   p   p   o   p   p  
380 o   p   p   p   o   p   p   o   p   p  
391 o   p   p   p   o   p   p   o   p   p  
394 n   n   o   o   o   o   n   o   o   o  
407 o   p   p   p   o   p   p   o   p   p  
414 o   p   p   p   o   p   p   o   p   p  
\end{Soutput}
\end{Schunk}

Strictly speaking, the diagonal of the balance semigroup structure with paths suggests that the system is not structurally balanced, and this is clearly due to the presence of the ambivalent tie, which has a devastating effect in the application of the algorithm. Nevertheless, ``weakly'' balanced structures may exist with classes of actors where the members of each class are all related through ambivalent ties \citep[see][for illustrative examples]{Ostoic17}.

Using the stability principle, we obtain a steady balance semiring structure with semipaths of length 3 where all the entries in the diagonal have a positive sign. This means that this component fulfills the requirements of the Structure Theorem with two \emph{groups} of actors having within positive ties and between negative connections and hence the system has an inherent equilibrium in it.

\begin{Schunk}
\begin{Sinput}
R> semiring(netCsg, type = "balance", k = 3)$Q
\end{Sinput}
\begin{Soutput}[fontsize=\small]
    328 342 352 368 376 380 391 394 407 414
328 p   p   p   p   p   p   p   n   p   p  
342 p   p   p   p   p   p   p   n   p   p  
352 p   p   p   p   p   p   p   n   p   p  
368 p   p   p   p   p   p   p   n   p   p  
376 p   p   p   p   p   p   p   n   p   p  
380 p   p   p   p   p   p   p   n   p   p  
391 p   p   p   p   p   p   p   n   p   p  
394 n   n   n   n   n   n   n   p   n   n  
407 p   p   p   p   p   p   p   n   p   p  
414 p   p   p   p   p   p   p   n   p   p  
\end{Soutput}
\end{Schunk}

The fact that the signed structure is structurally balanced implies that it is not prone to change from endogenous causes. However, the substantial implications of this resilience to change in a near future will depend on the relational content of the ties and in the context where the network of relations is taking place.

\subsubsection{Cluster semiring}
Another possibility of function \code{semiring} is to apply the \code{"cluster"} \code{type} option. The difference between this choice and balance semiring lies on the operation rules; cluster semiring namely handles as well an intermediate valence \code{q} that is the product of two negative signs \citep[cf.][and \citet{DorBatFer2004} for rules of cluster and balance semirings]{HarNorCat1965}. 

We evaluate next the signed network \code{netCsg} with the rules of the cluster semiring in order to obtain a stable arrangement, and by ``stable'' we mean that the semiring structure remains invariable with longer paths or chain relations. For instance, we verify that this algebraic structure considering paths with $k=4$ and semipaths with $k=5$ are identical, and hence we obtain the ending configuration to assess the network in terms of structural balance.

\begin{Schunk}
\begin{Sinput}
R> identical(
+    semiring(netCsg, type = "cluster", k = 4, symclos = FALSE)$Q,
+    semiring(netCsg, type = "cluster", k = 5, symclos = TRUE)$Q )
\end{Sinput}
\begin{Soutput}[fontsize=\small]
[1] TRUE
\end{Soutput}
\end{Schunk}

As a result, the steady cluster semiring structure for this component is obtained with paths of length 4, and this is because systems product of cluster semiring with longer paths remain unchanged. In this case, the constant structure distinguishes a single actor from the system, namely actor \textsf{394} who is only connected through negative ties to the rest of the system. However, since \code{a} acts as the absorbing element within the addition operation both with the balance and the cluster semiring versions, this actor is ambivalent tied from the rest of the component either with the path or the semipath options. The fact that all incoming chains of ties in the semiring structure are ambivalent, except for the cycle and semicycle involving actor \textsf{394}, suggests that the large component of Incubator network C is not structurally balanced as a clusterable system.

\begin{Schunk}
\begin{Sinput}
R> semiring(netCsg, type = "cluster", k = 4, symclos = FALSE)$Q
\end{Sinput}
\begin{Soutput}[fontsize=\small]
    328 342 352 368 376 380 391 394 407 414
328 a   a   a   a   a   a   a   a   a   a  
342 a   a   a   a   a   a   a   a   a   a  
352 a   a   a   a   a   a   a   a   a   a  
368 a   a   a   a   a   a   a   a   a   a  
376 a   a   a   a   a   a   a   a   a   a  
380 a   a   a   a   a   a   a   a   a   a  
391 a   a   a   a   a   a   a   a   a   a  
394 n   n   n   n   n   n   n   n   n   n  
407 a   a   a   a   a   a   a   a   a   a  
414 a   a   a   a   a   a   a   a   a   a  
\end{Soutput}
\end{Schunk}

The premise is that structurally balanced settings are expected to be more stable over time than imbalanced structures, and this allows supporting predictions e.g., about social influence processes through interpersonal relations. The presence of ambivalent ties prevent to have a structurally balanced structures, but it is still possible to find ``factions'' in a weakly balanced network of contrasting relations.


\section[Affiliation networks]{Affiliation networks}

Social systems in which the ``domain'' and ``co-domain'' do not coincide constitute \emph{affiliation networks}, which are also known as \emph{two-mode} or \emph{bipartite} networks, and this definition contrasts with the usual simple networks that are one-mode and in which there is just one set of relations on a single domain of nodes or social actors. \emph{Bipartite graphs} are naturally the representation form of bipartite or affiliation networks where the distinctive of the graph is that nodes are differentiated in two modes.

Affiliation networks are then systems of relations between two sets of entities that are naturally represented in \proglang{R} through data frames where intuitively rows and columns stand for the actors and their affiliations or attributes, respectively. 
For instance, object \code{G20} below records the memberships of the G20 countries according to \citet{wiki17} as a \code{data.frame} data object, where the supranational organizations are the column names. Later are specified the actors that are countries given in ISO 3166-1 alpha-3 codes in the row names.\footnote{We notice that the network here has 19 actor members, and this is because for practical reasons the European Union and other affiliated countries are disregarded.}

\begin{Sinput}
R> G20 <- data.frame( 
+    P5    = c(0,0,0,0,1,1,0,0,0,0,0,0,0,1,0,0,0,1,1),
+    G4    = c(0,0,1,0,0,0,1,1,0,0,1,0,0,0,0,0,0,0,0),
+    G7    = c(0,0,0,1,0,1,1,0,0,1,1,0,0,0,0,0,0,1,1),
+    BRICS = c(0,0,1,0,1,0,0,1,0,0,0,0,0,1,0,1,0,0,0),
+    MITKA = c(0,1,0,0,0,0,0,0,1,0,0,1,1,0,0,0,1,0,0),
+    DAC   = c(0,1,0,1,0,1,1,0,0,1,1,1,0,0,0,0,0,1,1),
+    OECD  = c(0,1,0,1,0,1,1,0,0,1,1,1,1,0,0,0,1,1,1),
+    Cwth  = c(0,1,0,1,0,0,0,1,0,0,0,0,0,0,0,1,0,1,0),
+    N11   = c(0,0,0,0,0,0,0,0,1,0,0,1,1,0,0,0,1,0,0) )

R> rownames(G20) <- c("ARG","AUS","BRA","CAN","CHN","FRA","DEU","IND","IDN",
+    "ITA","JPN","KOR","MEX","RUS","SAU","ZAF","TUR","GBR","USA")
\end{Sinput}

It is certainly possible to add different types of actor attributes in the network co-domain, but we are going to use such information in a different manner. For example, the ``economic classification'' of the countries according to the International Monetary Fund IMF is recorded as vector in \code{ac} below; first as a numeric vector, and then transformed into a character format. On the other hand, the clustering information for the ``events'' is given before in \code{ec}, and it encodes the supranational organizations that are ``bridges'' from the two kinds of ``assemblies'' representing block interests.

\begin{Schunk}
\begin{Sinput}
R> ec <- c(1,1,2,0,1,2,1,1,1)

R> ac <- c(0,1,0,1,0,1,1,0,0,1,1,1,0,0,0,0,0,1,1)
R> ac <- replace(ac, ac == 0, "Emerging")
R> ac <- replace(ac, ac == 1, "Advanced")
\end{Sinput}
\begin{Soutput}[fontsize=\small]
 [1] "Emerging" "Advanced" "Emerging" "Advanced" "Emerging" "Advanced" "Advanced"
 [8] "Emerging" "Emerging" "Advanced" "Advanced" "Advanced" "Emerging" "Emerging"
[15] "Emerging" "Emerging" "Emerging" "Advanced" "Advanced"
\end{Soutput}
\end{Schunk}

\subsubsection{Visualization of two-mode data}
\begin{figure*}[t!]
\vspace{-20pt}
\centering
\begin{tabular}{c}
\multicolumn{1}{c}{\includegraphics[width=11cm, page=7]{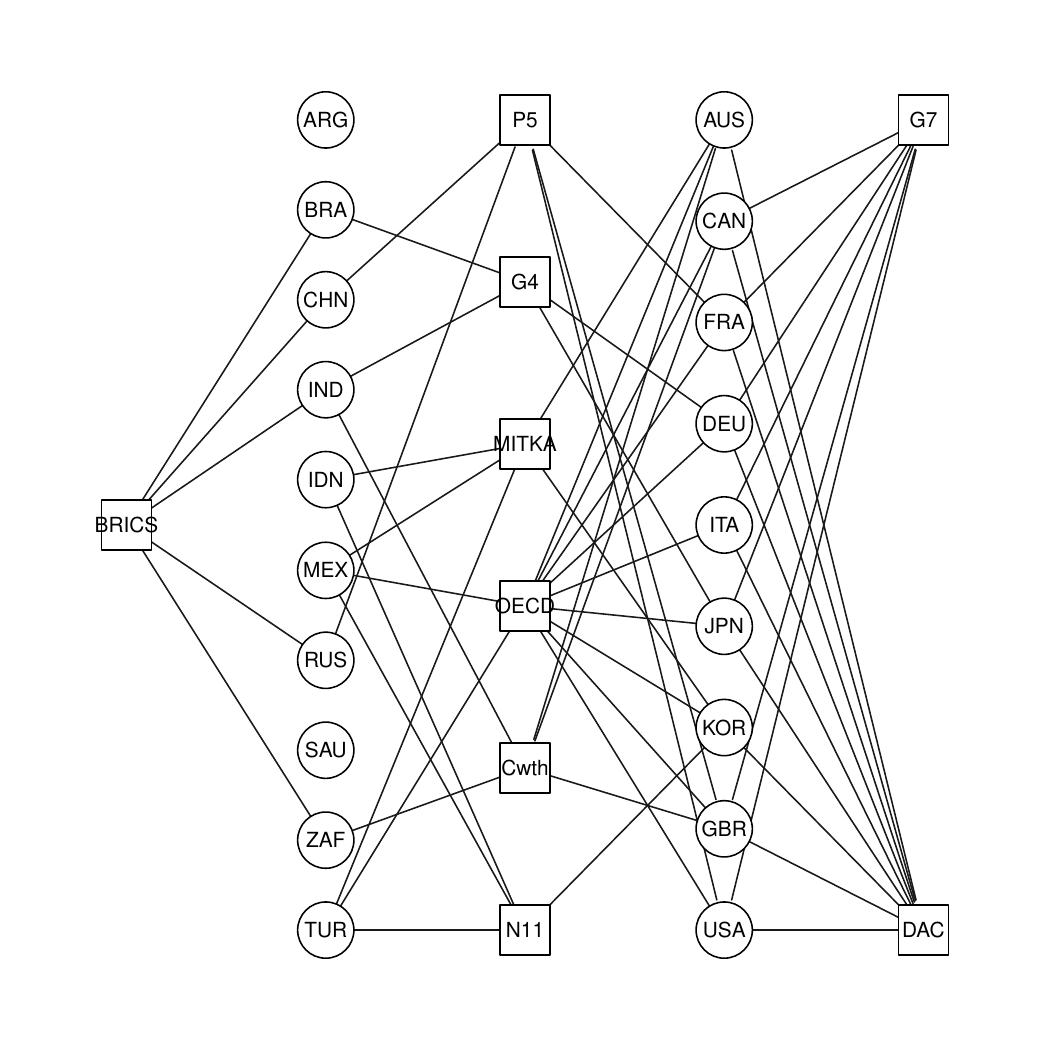} }\\[-110pt]
\multicolumn{1}{c}{\includegraphics[scale=1.2, page=8]{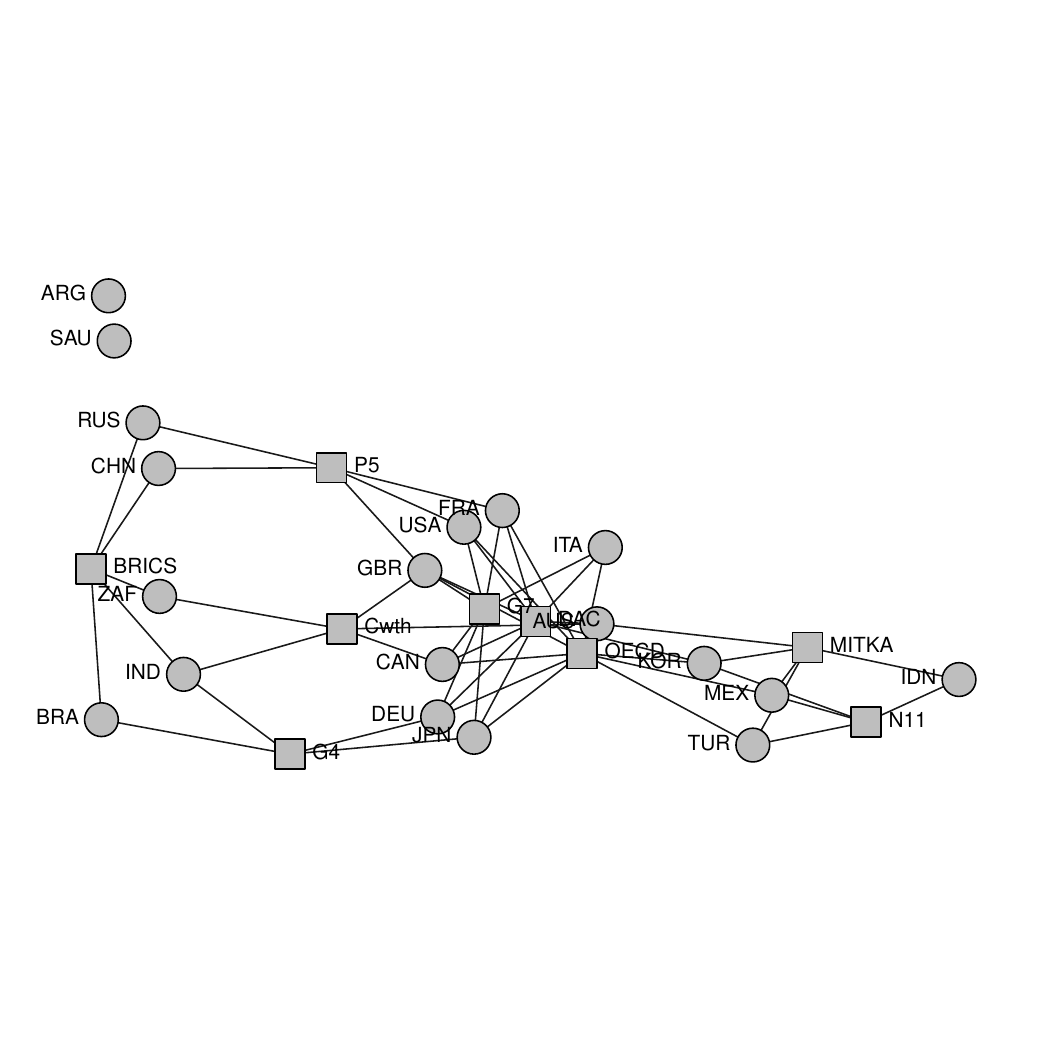} } 
\end{tabular}
\setlength{\abovecaptionskip}{-60pt}
  \caption{Clustered bipartite graph and a force-directed layout of the G20 affiliation network.}
\label{fig:G20bip}
\end{figure*}

The clustering information in vectors \code{ec} and \code{ac} is then used in the visualization of the \emph{bipartite graph}, and for this purpose, they need to be specified as a \code{list} object. 
Typically, bipartite graphs have two columns, one for each domain of the two-mode network; however, other options for the visualization are certainly possible. For instance, the \pkg{multigraph} function \code{bmgraph} with the code below plots in Figure~\ref{fig:G20bip} a bipartite graph with three columns where the actors are separated in different columns according to the clustering information given in \code{cluc} argument and the \code{layout} option \code{"bipc"}. On the other hand, the last command plots in the same Figure the network as a graph with a forced directed algorithm using the binomial projection approach to two-mode data described in \citet{Borgatti2012}.

\begin{Sinput}
R> bmgraph(G20, layout = "bipc", clu = list(ac, ec), cex = 3)
R> bmgraph(G20, layout = "force", seed = 321, vcol = 8, cex = 3, fsize = 8)
\end{Sinput}

The visualization of the data is an important step in the analysis of two-mode networks and both kinds of graphs show the countries distinct affinities with their memberships in their own manner. For instance, the force-directed layout clearly evidences some organizations that act as significant bridges between emergent and advances economies, while a bipartite graph typically relies on the given order of the elements in the object representing the network. In this case, however, there is a manual specification of the clustering, and it is worth mentioning that an increasing in the number of classes either in the domain or in the co-domain of the network implies an increasing in the number of columns in the graph with the \code{"bipc"} \code{layout} option of \code{bmgraph}.

\subsection{Formal concept analysis}
An algebraic approach for the analysis of affiliation networks is found in \emph{formal concept analysis} \citep{GantWill1996}. In terms of this analytical framework, the domain and co-domain of an affiliation network, respectively, are characterized as a set of \emph{objects} $G$ and a set of \emph{attributes} $M$. A formal ``context'' is obtained with an incidence relation $I \subseteq G \times M$ between these sets, which is the triple represented by the cross table of the data frame. 
The formal ``concept'' of a formal context is a pair of sets of objects $A$ and attributes $B$ that is maximally contained on each other; i.e., columns in the cross table representing the attribute set that help to cover the most entries in $I$, where $A$ and $B$ are said to be the ``extent'' and ``intent'' of the formal concept respectively.

A \emph{Galois derivation} between the power sets of sets $G$ and $M$ is derived for any subsets $A \subseteq G$ and $B \subseteq M$ by the set of attributes common to all the objects in the intent, $A^\prime$, and the set of objects possessing the attributes in the extent, $B^\prime$:

\begin{equation*}
\begin{aligned}
A^\prime \;=\; { m \in M \;\mid\; (g, m) \in I \quad(\text{for all } g \in A) } \\
B^\prime \;=\; { g \in G \;\mid\; (g, m) \in I \quad(\text{for all } m \in B) }
\end{aligned}
\end{equation*}
where $g$ and $m$ are rows and columns in the derivation operation, and $G$ and $M$ result being two closed systems dually isomorphic to each other.

With \pkg{multiplex}, it is possible to obtain the Galois derivations between the objects and the attributes in a given context with the \code{galois} function that produces a \code{"Galois"} class object. For instance, a truncated output of these connections in the \code{G20} data set is given below with the default \code{"full"} \code{labeling} argument of the elements in this formal context, and there are different sets of $G$ elements in this class object recorded as a list data type where diverse sets of $M$ are the attribute names of the items. We can see, for example, that this part of the output shows affiliations of the three organizations that are interest assemblies in the network and three bridge organizations. Moreover, the \textsf{G7} countries are also members of \textsf{DAC} and \textsf{OECD}, and some of these countries are connected to the organizations previously derived.

\begin{Schunk}
\begin{Sinput}
R> galois(G20)
\end{Sinput}
\begin{Soutput}[fontsize=\small]
$P5
[1] "CHN, FRA, GBR, RUS, USA"

$G4
[1] "BRA, DEU, IND, JPN"

$`DAC, G7, OECD`
[1] "CAN, DEU, FRA, GBR, ITA, JPN, USA"

$BRICS
[1] "BRA, CHN, IND, RUS, ZAF"
...
\end{Soutput}
\end{Schunk}

Such output is not simply a set of elements, but actually constitutes a \emph{family} of concepts where the order of the elements is significant. The \code{galois} function allows as well to obtain a \code{"reduced"} \code{labeling} of objects and attributes, which in most of the cases provides a more informative output. With this option, the repeated objects discard the previous ones, whereas the recurrent attributes in the listing are also discarded but afterward in the derivation. 

The Galois connections of network \code{G20} is given below with the reduced labeling, and such output is recorded in object \code{G20gc} for a later use. We bear in mind that the order of the concepts is the same as with the full labeling given earlier, and we evidence, for example, that in the third concept there is a single country that is related just to the three attributes of the full labeling without belonging to any other concept. However, two of the three organizations in this concept have other members (not printed here) and therefore their labels are located in concepts listed afterward the derivation.

\begin{Schunk}
\begin{Sinput}
R> G20gc <- galois(G20, labeling = "reduced")
\end{Sinput}
\begin{Soutput}[fontsize=\small]
$reduc
$reduc$P5
character(0)

$reduc$G4
character(0)

$reduc$G7
[1] "ITA"

$reduc$BRICS
character(0)
...
\end{Soutput}
\end{Schunk}

\subsubsection{Partial ordering of the concepts}
In the previous example, recurrent objects in the family of concepts with reduced labeling discard the previous ones produced in the derivation. This implies that these latter objects are covered for the one printed in the output, and such type of relation applies to all objects, and dually to the attributes as well. Hence, the set of inclusions of the concepts serves to clarify the disposition of the Galois derivations in the formal context.

The \emph{partial ordering of the concepts} is established as the hierarchy given by the relation subconcept--superconcept, $(A_1, B_1)$ and $(A_2, B_2)$, of extents and intents, which is formally expressed as:

$$
(A, B) \;\leq\; (A_2, B_2) \quad\Leftrightarrow\quad A_1 \subseteq A_2 \qquad (\;\Leftrightarrow\quad B_1 \;\subseteq\; B_2\;).
$$

The set of formal concepts product of the Galois derivations in the formal context serves to establish the partial ordering of the concepts. For the G20 countries formal context, the ordering among the concepts is given in \code{G20gc}, which contains the \emph{full} derivation of the objects and attributes even if is specified with a \code{"reduced"} labeling in the \code{galois} function.

To obtain the ordered structure of the Galois derivations of the formal concepts, we use the \code{"galois"} \code{type} argument in the function \code{partial.order}, which produces a \code{"Partial.Order"} class object that is a ``galois'' type as well. As said, although the reduced option is typically used for the representation of the inclusion structure, the hierarchy given in the partially ordered structure is based on the  ``full'' labeling of the concepts is part of the output even though it is not printed on the screen with the reduced option. To have a better representation of the partial order structure, the \code{lbs} argument in the function allows customizing the names of the concepts in the context corresponding to the matrix labels.

\begin{Schunk}
\begin{Sinput}
R> partial.order(G20gc, type = "galois", lbs = paste0("c", seq(1, 25)))
\end{Sinput}
\begin{Soutput}[fontsize=\footnotesize]
    c1 c2 c3 c4 c5 c6 c7 c8 c9 c10 c11 c12 c13 c14 c15 c16 c17 c18 c19 c20 c21 c22 c23 c24 c25
c1   1  0  0  0  0  0  0  0  0   0   0   0   0   0   0   0   0   0   0   0   0   0   0   0   1
c2   0  1  0  0  0  0  0  0  0   0   0   0   0   0   0   0   0   0   0   0   0   0   0   0   1
c3   0  0  1  0  0  1  1  0  0   0   0   0   0   0   0   0   0   0   0   0   0   0   0   0   1
c4   0  0  0  1  0  0  0  0  0   0   0   0   0   0   0   0   0   0   0   0   0   0   0   0   1
c5   0  0  0  0  1  0  0  0  0   0   0   0   0   0   0   0   0   0   0   0   0   0   0   0   1
c6   0  0  0  0  0  1  1  0  0   0   0   0   0   0   0   0   0   0   0   0   0   0   0   0   1
c7   0  0  0  0  0  0  1  0  0   0   0   0   0   0   0   0   0   0   0   0   0   0   0   0   1
c8   0  0  0  0  0  0  0  1  0   0   0   0   0   0   0   0   0   0   0   0   0   0   0   0   1
c9   0  0  0  0  1  0  0  0  1   0   0   0   0   0   0   0   0   0   0   0   0   0   0   0   1
c10  1  1  1  1  1  1  1  1  1   1   1   1   1   1   1   1   1   1   1   1   1   1   1   1   1
c11  1  0  1  0  0  1  1  0  0   0   1   0   0   0   0   0   0   0   0   0   0   0   0   0   1
c12  1  0  0  1  0  0  0  0  0   0   0   1   0   0   0   0   0   0   0   0   0   0   0   0   1
c13  1  0  1  0  0  1  1  1  0   0   1   0   1   0   0   0   1   0   0   0   0   1   0   0   1
c14  0  1  1  0  0  1  1  0  0   0   0   0   0   1   0   0   0   0   0   0   0   0   0   0   1
c15  0  1  0  1  0  0  0  0  0   0   0   0   0   0   1   0   0   0   0   0   0   0   0   0   1
c16  0  1  0  1  0  0  0  1  0   0   0   0   0   0   1   1   0   1   0   0   0   0   0   0   1
c17  0  0  1  0  0  1  1  1  0   0   0   0   0   0   0   0   1   0   0   0   0   1   0   0   1
c18  0  0  0  1  0  0  0  1  0   0   0   0   0   0   0   0   0   1   0   0   0   0   0   0   1
c19  0  0  0  0  1  1  1  0  0   0   0   0   0   0   0   0   0   0   1   1   0   0   0   0   1
c20  0  0  0  0  1  0  1  0  0   0   0   0   0   0   0   0   0   0   0   1   0   0   0   0   1
c21  0  0  0  0  1  1  1  1  0   0   0   0   0   0   0   0   0   0   1   1   1   1   0   0   1
c22  0  0  0  0  0  1  1  1  0   0   0   0   0   0   0   0   0   0   0   0   0   1   0   0   1
c23  0  0  0  0  1  1  1  0  1   0   0   0   0   0   0   0   0   0   1   1   0   0   1   1   1
c24  0  0  0  0  1  0  1  0  1   0   0   0   0   0   0   0   0   0   0   1   0   0   0   1   1
c25  0  0  0  0  0  0  0  0  0   0   0   0   0   0   0   0   0   0   0   0   0   0   0   0   1
attr(,"class")
[1] "Partial.Order" "galois"
\end{Soutput}
\end{Schunk}

\subsubsection{Concept lattice of the context}
By looking at the partial order table above, we can see for instance that concept \textsf{10} is contained by all elements of the structure, whereas concept \textsf{25} includes the rest of the concepts resulting from the Galois derivations. This means that the former concept constitutes the minimal element in the lattice and is covered by the rest of the concepts of the partial order structure, whereas the latter concept is the maximal element in the lattice and covers the rest of the elements in the poset. However, other inclusion relations may be difficult to observe with the matrix format and the visualization of the partially ordered structure of inclusions among formal concepts results most of times more informative.

The \emph{concept lattice} of the formal context, aka \emph{Galois lattice} in social network analysis, is established as a system of partially ordered elements corresponding to the set of all concepts. In this case, the greatest lower bound of the meet and the least upper bound of the join are defined in terms of objects and attributes and an index set $T$ by the \emph{Basic Theorem of Concepts Lattices} \citep{GantWill1996}.

\begin{equation*}
\begin{aligned}
\bigwedge_{t\in T} \quad \bigl(A_t, B_t \bigr) \;=\; \Bigl(\; \bigcap_{t\in T}{A_t}, \;\;\bigl(\bigcup_{t\in T}{B_t} \bigr)^{\prime\prime} \;\Bigr)\phantom{.} \\
\bigvee_{t\in T} \quad \bigl(A_t, B_t \bigr) \;=\; \Bigl(\; \bigl(\bigcup_{t\in T}{A_t} \bigr)^{\prime\prime}, \;\;\bigcap_{t\in T}{B_t} \;\Bigr). \\
\end{aligned}
\end{equation*}

To create the partial order structure of the Galois connections among the concepts in formal context of the G20 affiliation network, we use the \code{partial.order} function with the appropriate \code{type}. The \code{"dimnames"} of the poset structure in object \code{G20po} reflects the labeling of the derivations used in the input, which in this case was specified with the reduced labeling, and the construction of the concept lattice through the \code{diagram} function lies on this information.

\begin{Sinput}
R> G20po <- partial.order(G20gc, type = "galois")
R> diagram(G20po)
\end{Sinput}

\begin{figure*}[t]
\vspace{-25pt}
    \centering\includegraphics[width=16cm,page=9]{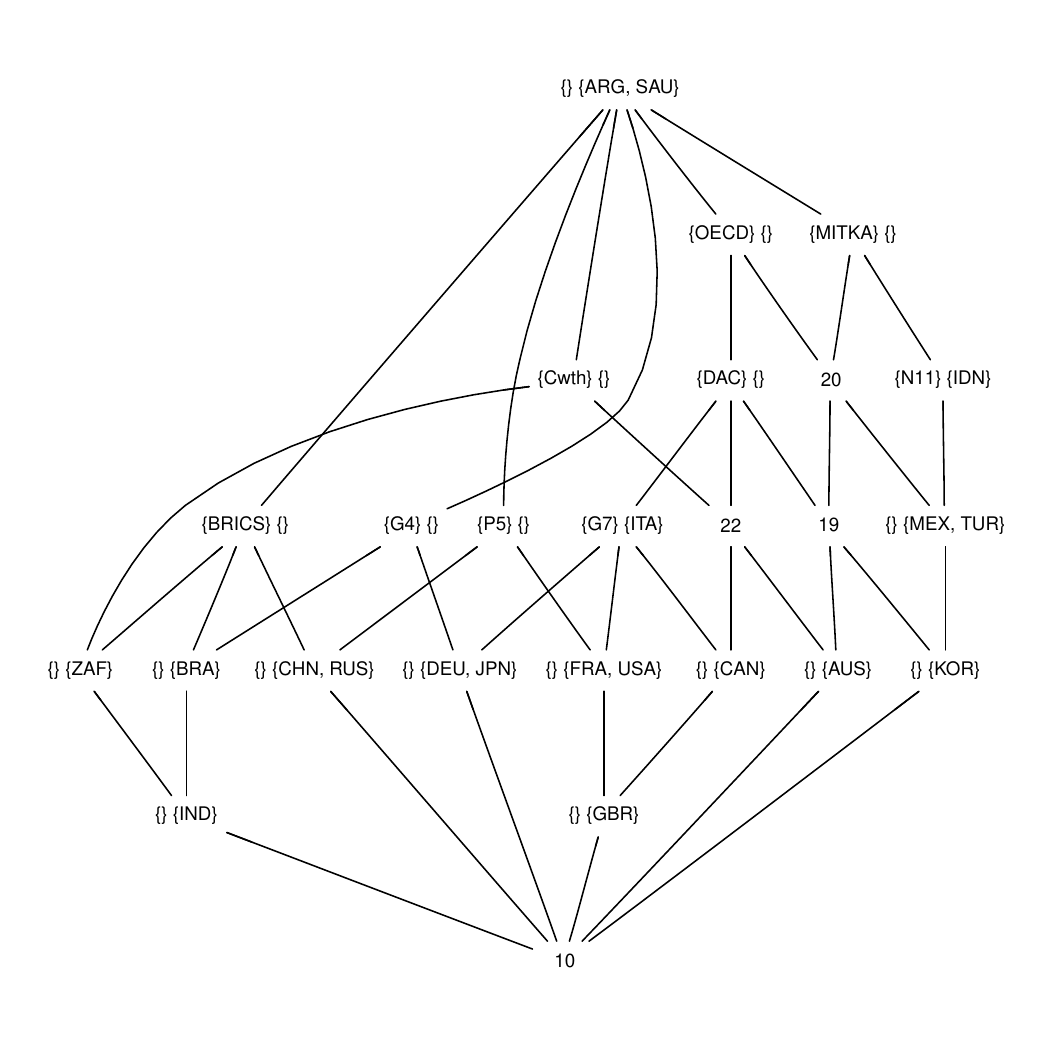}
\setlength{\abovecaptionskip}{-25pt}
  \caption{Concept lattice of the G20 countries and their affiliations.}
\label{galois}
\end{figure*}

Figure~\ref{galois} depicts as a concept lattice the set of inclusions among the maximal rectangles in the formal context corresponding to the G20 countries and their affiliations and, since this is a reduced representation of the context, both objects and attributes are given just once as in the partial order structure. We notice that ---contrary to the typical representation of a formal context--- the labels of the concepts are placed instead of the nodes rather than next to them. When a concept does not have a label, which most of the times occurs in reduced contexts, then the respective number of the concept is placed instead the node rather than leaving it unlabeled as it happens with concepts \textsf{10}, \textsf{19}, \textsf{20}, and \textsf{22}.

In the concept lattice of the context, objects having more attributes are located downward and hence covered by the objects with less attributes; conversely, the reverse is true for the attributes, which means that the most popular ones are located more upward than the less popular attributes. This implies that the ``levels'' in the lattice somewhat tries to reflect the covering in the two instances. However, the placement of the elements in the lattice diagram corresponds to the layout algorithm applied by the \pkg{Rgraphviz} package, which can result rather arbitrary depending on the cases. In this case, although \textsf{ZAF} has fewer affiliations than \textsf{ITA} for instance, it is depicted in a lower level than the latter country (cf. Fig.~\ref{galois}), and to have a more precise information we need to count with the set of inclusion relations in the context lattice, particularly in lattice structures with a large number of concepts.

\subsection{Order filters and order ideals}
The construction of the concept lattice of the context allows us to gain significant information about the affiliation network, and one part of the lattice structure is concerned with the inclusion levels among the concepts. Another aspect deals with ``downsets'' and ``upsets'' that are implications made from all the lower and greater bounds of a given element in the line diagram. For relatively large formal contexts, however, the corresponding poset structure typically results in a too complex, and it is useful to count with the set of implications among objects and attributes in the concept lattice structure of any given formal context. Algebraic notions based on upsets and downsets are of fundamental importance in uncovering implication subsets in an ordered structure, and next we look at requires a formal definition of these theoretical concepts. 

An ordered set is represented by the pair $(P, \quad \leq)$ where $a$ and $b$ are some elements in $P$. A non-empty subset $U$ [resp. $D$] of $P$ is an upset [resp. downset] called a \emph{order filter} [resp. \emph{order ideal}] iff, for all $a \in P$ and $b \in U$ [resp. $D$]:
$$
b \leq a \text{\quad implies\quad} a \in U  \qquad\qquad [\;\text{resp.\;\;} a \leq b \text{\quad implies\quad} a \in D\;]
$$

For a specific element $x \in P$, the upset $\uparrow\! x$ formed for all the upper bounds of $x$ is called a \emph{principal order filter} generated by $x$. Dually, $\downarrow\! x$ is a \emph{principal order ideal} with all the lower bounds of $x$. Order filters and order ideals not coinciding with $P$ are called \emph{proper}. In this sense, both order filters and order ideals of particular elements of the context are based on the partial ordering of the concepts. 

Function \code{fltr} of \pkg{multiplex} allows finding these subsets of elements in the context lattice either with the label or with the number of the concept, which has been assigned according to the corresponding partial order structure. 
For example, the principal order filter set of the third element in the partial order structure of the Galois derivations of the concepts corresponding to the G20 affiliation network, which is obtained with the default option in this function. 
The output starts with the concept used as the input, and then continues with the different inclusions to this concept until the maximal element, and the numbering of the concepts according to the partial order table corresponds to the attribute names of the items of the list data type of the output.

\begin{Schunk}
\begin{Sinput}
R> fltr(3, G20po)
\end{Sinput}
\begin{Soutput}[fontsize=\small]
$`3`
[1] "{G7} {ITA}"

$`6`
[1] "{DAC} {}"

$`7`
[1] "{OECD} {}"

$`25`
[1] "{} {ARG, SAU}"
\end{Soutput}
\end{Schunk}

Likewise, proper order ideals of one or more concepts are obtained with the same function provided that the \code{ideal} argument is activated in the input. In the next example, we employ the name of the intents of two concepts, which gives the affiliates of \textsf{G7} and \textsf{BRICS} in the G20 network.

\begin{Schunk}
\begin{Sinput}
R> fltr(c("G7" , "BRICS"), G20po, ideal = TRUE)
\end{Sinput}
\begin{Soutput}[fontsize=\small]
$`3`
[1] "{G7} {ITA}"

$`4`
[1] "{BRICS} {}"

$`10`
[1] "10"

$`11`
[1] "{} {FRA, USA}"

$`12`
[1] "{} {CHN, RUS}"

$`13`
[1] "{} {GBR}"

$`14`
[1] "{} {DEU, JPN}"

$`15`
[1] "{} {BRA}"

$`16`
[1] "{} {IND}"

$`17`
[1] "{} {CAN}"

$`18`
[1] "{} {ZAF}"
\end{Soutput}
\end{Schunk}

Although in this case the membership to the two organizations may be known in advance, order filters and order ideals are convenient ways to discover recurring sets of dependences amongst different types of formal concepts. Particularly for objects and attributes belonging to large affiliation networks and other complex systems with separate domains.


\section[Discussion]{Discussion}

\pkg{multiplex} is one of the first attempts to provide algebraic procedures for the analysis of complex network structures within the \proglang{R} environment, and this is despite the methodologies have been developed for some years ago. Algebraic analysis is characterized by its deterministic reasoning, and this constitutes a significant difference with the statistical approach. One advantage of statistics over algebra is that it can handle large amounts of information, whereas the benefit from an algebraic analysis of networks is that we have the certainty that the resulted structure is taken place somewhere in the system, and it is not merely a likelihood as with statistical inference.

Algebraic systems such as semigroups and semirings are aimed to relative small to medium size network structures, which constitutes a significant limitation in some cases. This also applies to Compositional equivalence since there is the risk that the accumulation of perceived hierarchies in the network ends up having no structure as the universal matrix, and instead of individuals, semigroups for instance are typically applied to role structures made of categories of actors. For larger networks, such categories may be produced with a model-fitting approach where a number of classes of hierarchical exponential-family random graph models for social networks constitute an alternative to the descriptive and are able to handle multiplex networks. A crucial step to obtain the logic of interlock in the network and algebraic constraints governing the relations among relations (such as set of equations, hierarchy of string relations, and role tables) is , however, the decomposition of the semigroup structure. \pkg{multiplex} can handle both abstract and partially ordered semigroups for decomposition.

With respect to algebraic approaches for signed structures, there is also a concern with the network size. Working with categories of actors is, again, a possibility for large networks, but this still is today a relative unexplored area for a potential research. One aspect worth to mention is that one can see balance and cluster semirings as fuzzy structures where the positive, negative, and ambivalent valences constitute the values in the fuzzy set, which can even be continuous in case the reduction of the valence structure produces in the image matrix relations with mixed signs among factions of differentiated actors.

Visualizing poset structures is especially convenient for deducing containments among elements in the partial ordering, and this applies both to partially ordered semigroups and to concept lattices of a given formal context. It is a very difficult task ---if not impossible--- just by looking at the array representing hierarchies of either string relations of multiplex networks, or of Galois derivations of formal concepts corresponding to affiliation networks. For large partially ordered structures, it may be necessary to apply the algebraic notions of order filters and order ideals to produce particular sets of inclusions, and this makes possible to elaborate substantial interpretations of the hierarchy of concepts in a systematic way, as the set of equations and role tables do for the relational interlock of multiplex network structures.

If possible, stochastic and algebraic analyzes should complement to each other in algebraic statistics, and the condensation of large network structures is likely to be made in statistical terms, leaving the algebra to make the more subtle examination of the reduced structure. In this sense, the integration of statistical and algebraic approaches constitutes a promising line of research within the structural analysis, and \proglang{R} provides an adequate setting to study and test novel structural methods and theoretical models with real world applications.


\bibliography{./JSSarXiv}

\end{document}